\documentclass[usenatbib]{mnras}
\usepackage{aas_macros}
\usepackage{natbib}
\usepackage[utf8]{inputenc}
\usepackage{amsmath}
\usepackage{amssymb}
\usepackage{graphicx}
\usepackage[dvipsnames]{xcolor}
\hypersetup{colorlinks=true,allcolors=teal}
\usepackage[normalem]{ulem}
\usepackage{fontawesome}

\newcommand{\p}{\ensuremath{\partial}}
\newcommand{\Msun}{\ensuremath{M_{\odot}}}

\newcommand{\Mpch}{\ensuremath{h^{-1}{\rm Mpc}}}
\newcommand{\Mpcinvh}{\ensuremath{h{\rm Mpc^{-1}}}}

\newcommand{\avg}[1]{\ensuremath{\left\langle \,#1\, \right\rangle}}
\newcommand{\e}[1]{\ensuremath{{\rm e}^{#1}}}

\newcommand{\der}{\ensuremath{{\rm d}}}

\newcommand{\eqn}[1]{equation~\eqref{#1}}
\newcommand{\eqns}[1]{equations~\eqref{#1}}

\newcommand{\be}{\begin{equation}}
\newcommand{\ee}{\end{equation}}

\title[Separate universe bias \& tidal anisotropy]
      {Separate Universe calibration of the dependence of halo bias on cosmic web anisotropy} 
\author[Ramakrishnan \& Paranjape]{
Sujatha Ramakrishnan$^{1}$\thanks{E-mail: rsujatha@iucaa.in}, 
Aseem Paranjape$^{1}$\thanks{E-mail: aseem@iucaa.in}
\\  
 $^1$ Inter-University Centre for Astronomy \& Astrophysics,
      Ganeshkhind, Post Bag 4, Pune 411007, India}
\date{draft}

\begin{document}
\label{firstpage}
\pagerange{\pageref{firstpage}--\pageref{lastpage}}
\maketitle

\begin{abstract}
We use the Separate Universe technique to calibrate the dependence of linear and quadratic halo bias $b_1$ and $b_2$ on the local cosmic web environment of dark matter haloes. 
We do this by measuring the response of halo abundances at fixed mass and cosmic web tidal anisotropy $\alpha$ to an infinite wavelength initial perturbation. 
We augment our measurements with an analytical framework developed in earlier work which exploits the near-Lognormal shape of the distribution of $\alpha$ and results in very high precision calibrations.  
We present convenient fitting functions for the dependence of $b_1$ and $b_2$ on $\alpha$ over a wide range of halo mass for redshifts $0\leq z\leq1$. 
Our calibration of $b_2(\alpha)$ is the first demonstration to date of the dependence of non-linear bias on the local web environment. 
Motivated by previous results which showed that $\alpha$ is the primary indicator of halo assembly bias for a number of halo properties beyond halo mass, we then extend our analytical framework to accommodate the dependence of $b_1$ and $b_2$ on any such secondary property which has, or can be monotonically transformed to have, a Gaussian distribution. 
We demonstrate this technique for the specific case of halo concentration, finding good agreement with previous results.  
Our calibrations will be useful for a variety of halo model analyses focusing on galaxy assembly bias, as well as analytical forecasts of the potential for using $\alpha$ as a segregating variable in multi-tracer analyses. 
\end{abstract}


\section{Introduction}
The large-scale clustering of gravitationally bound haloes of dark matter is a key variable in understanding the formation and evolution of the large-scale structure of the Universe \citep[see][for a review]{djs18}. This `halo bias' is known to depend on a number of halo properties such as halo mass \citep{Kaiser84,bbks86,mw96,st99}, halo assembly history \citep{st04,gsw05,wechsler+06}, halo shape, angular momentum and kinematics \citep{fw10} and the local tidal environment \citep{sams06,hahn+09,bprg17,phs18a,rphs19}. The dependence of halo bias on secondary properties beyond halo mass, generically referred to as `halo assembly/secondary bias', has emerged as a robust prediction of the hierarchical $\Lambda$-cold dark matter ($\Lambda$CDM) structure formation paradigm. Typically, halo assembly bias in some halo property $c$ (such as concentration, age, spin, ellipticity, velocity anisotropy, etc.) manifests as a difference in mean bias, at fixed halo mass, between halo populations having large and small values of $c$. Although there has been some analytical progress in describing such trends using simplified models 
(see, e.g., \citealp{zentner07,dwbs08,desjacques:2008a,ms12,cs13}
; \citealp{musso+18}), many of these trends show complex behavior, e.g. when multiple secondary variables are studied simultaneously (\citealp{lms17}; \citealp{mzw18}; \citealp{xz18,han+19}). A detailed understanding of halo assembly bias from first principles is therefore currently an open problem.

On another front, if the physics of galaxy formation and evolution couples tightly to the mass accretion history of dark matter haloes \citep{wr78} -- as is routinely assumed in semi-analytic models (SAMs) of galaxy evolution \citep[e.g.,][]{henriques+15} as well as (sub)-halo abundance matching (SHAM) exercises \citep{rwtb13,hw13,zehavi+19,caz20} and also confirmed by cosmological hydrodynamical simulations \citep{chaves-montero+16,bray+16,montero-dorta+20} -- then one expects \emph{galaxy} assembly bias trends to be apparent in observed galaxy samples. Due to systematic uncertainties in cleanly segregating observed samples, however, such trends have been difficult to establish robustly, with many conflicting results (\citealp{lin+16,miyatake+16}; \citealp{more16}; \citealp{zentner+16,montero-dorta+17,tinker+17,zu+17,tojeiro+17}; \citealp{busch17,obuljen20}). A unified framework to understand halo and galaxy assembly bias is therefore currently lacking.

Some recent developments are noteworthy in this context. Studies using dark matter only $N$-body simulations have demonstrated that the \emph{local tidal environment} of haloes plays a key role in explaining many (if not most) of the halo assembly bias trends studied in the literature. The tidal environment of a halo can be conveniently quantified by the \emph{tidal anisotropy} $\alpha$ constructed using the tidal tensor of the cosmic web in the vicinity of the halo \cite[][see below for details]{phs18a}. This variable has been shown to have the strongest correlation with large-scale bias amongst a number of secondary halo properties, and also statistically explains the assembly bias of all these properties \citep{rphs19}. The origins of some of these correlations, such as those between $\alpha$ and the halo age, concentration and velocity anisotropy, can be understood in terms of the dynamics of mass accretion as revealed by using high-resolution zoom simulations of objects accreting in and outside cosmic filaments \citep{hahn+09,bprg17}. Although a complete dynamical understanding of all the correlations is lacking, it is still possible to use simulations to calibrate these correlations. 

Our focus in this work is the relation between tidal anisotropy $\alpha$ and the large-scale halo bias. The calibration of this relation at fixed halo mass is most efficiently done using the Separate Universe (henceforth, SU) technique \citep{tb96,cole97,bssz11,lht16} which provides an exact realization of the peak-background split \citep{lwbs16}. Moreover, when augmented by some basic analytical modeling of the statistical distribution of the underlying variables, the SU technique can provide unprecedented precision in calibrating secondary bias at fixed halo mass, as demonstrated by \citet{pp17} for halo concentration \citep[see also][]{lms17}. In this paper, we will use these tools to calibrate the relation between $\alpha$ and the linear ($b_1$) and quadratic ($b_2$) bias of dark matter haloes. This calibration then becomes potentially useful for a number of applications which require accurate modeling of correlations between large-scale clustering and small-scale halo properties, such as analytical halo models of assembly bias, generating mock halo catalogs with accurate halo assembly bias using low-resolution simulations, forecasting multi-tracer cosmological constraints, etc., some of which we will discuss below.

The paper is organized as follows. Section~\ref{sec:sim&halo_prop} describes the SU simulations and halo properties used in this work. In Section~\ref{sec:alpha_calibration}, we present our calibration of the dependence of $b_1$ and $b_2$ on the tidal anisotropy $\alpha$. In Section~\ref{sec:conc_ab}, we extend the analytical framework mentioned above to include the dependence on both $\alpha$ and a secondary variable $c$ in $b_1$ and $b_2$, focusing on halo concentration as a specific example. We conclude in Section~\ref{sec:summary}. The Appendices present some technical details and calculations relevant to the main text.

\section{Simulations and Halo Properties}
\label{sec:sim&halo_prop}

\subsection{Separate Universe simulations}
The peak-background split halo bias parameters are defined in terms of the derivative of the mean number density of haloes with respect to the infinite wavelength density perturbation, i.e., as response coefficients. The response of halo number density to the presence of such a perturbation in a local region of the fiducial FLRW universe is identical to that produced in a universe with a modified cosmology having a larger/smaller physical background density depending on the sign of the perturbation. If we denote the infinite wavelength perturbation linearly extrapolated to present day as $\delta_L$, then in practice the SU technique takes a fiducial universe with $\delta_L\neq 0$ and performs an exact mapping to a curved universe with a different spatial curvature, matter density parameter and Hubble constant, all determined by the value of $\delta_L$. We refer the reader to \citet{wsck15a} for details of the numerical implementation of the $\delta_L \rightarrow {\rm FLRW}$ mapping in $N$-body simulations.

In the following, we give a few details regarding the simulations, halo identification and cleaning procedure, which are identical to \citet{pp17}.\footnote{ \url{https://bitbucket.org/aparanjape/separateuniversescripts}} Hence we refer the reader to the same for a more elaborate discussion. For our fiducial cosmology, we use a flat $\Lambda$CDM model with total matter density parameter $\Omega_m = 0.276$, baryonic matter density parameter $\Omega_b=0.045$, Hubble constant $H_0=100h \rm kms^{-1} Mpc^{-1}$ with $h=0.7$, primordial scalar spectral index $n_{s}=0.961$ and amplitude of linear perturbations smoothed on a comoving scale $8\Mpch$ $\sigma_8=0.811$.
Our $N$-body simulations are performed using \textsc{gadget-2} \citep{springel:2005}\footnote{ \url{http://www.mpa-garching.mpg.de/gadget/}}.
All the simulations have a comoving box size $L_{box} = 300/0.7$  Mpc and a particle count of $512^3$ each. In addition to the fiducial cosmology, we use a set of simulations generated with the SU technique that correspond to $\delta_L\in$ $\{\pm0.7,$ $\pm0.5,$ $\pm0.4,$ $\pm0.3,$ $\pm0.2,$ $\pm0.1,$ $\pm0.07,$ $\pm0.05,$ $\pm0.02,$ $\pm0.01,$ $+0.15,$ $+0.25,$ $+0.35\}$.
Our SU simulations are performed keeping the comoving box size fixed at all redshifts \citep[see][for other variants]{wsck15a}. Since the physical matter density parameter $\Omega_m h^2$ is the same in all the boxes, the particle mass $m_{part} = 2.2 \times10^{10} M_{\odot}$ is also the same in all the simulations. We have 15 sets of simulations for each $\delta_L$ performed by changing the seed for the random initial conditions, while keeping the seed the same across all $\delta_L$ values in each set. Additionally,    10 realizations of higher resolution ($1024^3$ particles) $\delta_L=0$ boxes are also used in order to test for convergence of various quantities computed.

Haloes are identified using \textsc{rockstar} \citep{behroozi13-rockstar}\footnote{ \url{https://bitbucket.org/gfcstanford/rockstar}} which uses a 6-dimensional Friends-of-Friends algorithm to make catalogs of haloes and their properties. From the catalog, only host haloes are chosen so that the analysis is unaffected by substructure.
Haloes were chosen to have a  minimum of 400 particles inside the radius $R_{200b}$ (see below). Unrelaxed haloes with `virial ratio' $2T/|U|\geqslant2$ are removed from our analysis \citep[see][for a detailed discussion]{bett+07}.

 In the SU approach, the fiducial universe at redshift $z$ is mapped to a universe with a modified cosmology at ${\tilde z}$ and their background densities are related by
\be
\tilde{\varrho}(t) = \varrho(t) \dfrac{(1+{\tilde z})^3}{(1+z)^3}\,.
\ee
Here, the notations $\varrho(t)$ and $\tilde{\varrho}(t)$ are similar to \citet{wsck15a}\footnote{The redshifts $z$ and ${\tilde z}$ can be related by equating the cosmic age integrals $\int_{z}^{\infty}{\der z^\prime}/{[H_{\rm fid}(z^\prime)(1+z^\prime)]}=\int_{{\tilde z}}^{\infty}{\der{z^\prime}}/{[\tilde{H}({z^\prime})(1+{z^\prime})]}$ where $H_{\rm fid}$ and $\tilde{H}$ are the Hubble parameters corresponding to the fiducial and modified cosmologies, respectively.} and stand for  the physical background matter density of the fiducial and modified cosmology, respectively, at cosmic time $t$. Among many other halo properties, \textsc{rockstar} calculates a value of $M_{200b}$ for each halo, defined as the mass inside a sphere of radius $R_{200b}$ within which the average density of the halo is 200 times the background density of the universe. However, \textsc{rockstar} uses the modified background density rather than fiducial background density to compute quantities like $M_{200b}$. Hence we can ensure that we get $M_{200b}$ of our fiducial cosmology \citep{lwbs16} by configuring \textsc{rockstar} to output $M_{\Delta b}$ where $\Delta = 200 {(1+z)^3}/{(1+{\tilde z})^3}$. Throughout this paper, we will work with $M_{200b}$ to represent the mass $m$ of the halo.

\subsection{Halo bias with the SU technique}
\label{sec:computeb1}
The overdensity of haloes in a $\delta_L \ne 0$ Lagrangian  patch is given in terms of the differential number density $n(m,\delta_L)$ of haloes between masses $(m,m+\der m)$ as follows
\be
 \delta_{h}^{L}(m,\delta_L)\equiv \dfrac{n(m|\delta_L)}{n(m|\delta_{L}=0)}-1\,.
 \label{eq:del-num}
\ee
It can also be related to the underlying dark matter distribution in terms of bias coefficients $b^L_n(m)$ as
\be
 \delta_{h}^{L}(m,\delta_L) = \sum_{n=1}^{\infty}\dfrac{b_n^L(m)}{n!}\delta^n_L\,.
 \label{eq:del-delL}
\ee
Equating the right hand sides (RHS) of \eqns{eq:del-num} and \eqref{eq:del-delL} allows us to extract the bias coefficients from number density measurements as we describe next. We have several $\delta_L\ne 0$ simulation boxes, each having the same number of particles of identical mass; hence all the SU simulations have identical Lagrangian volume. Thus the numerator and denominator in \eqn{eq:del-num} can be replaced by the number count of haloes between mass $(m,m+\der m)$ in our simulation boxes. 
We compute the RHS of \eqn{eq:del-num} for each realization and average over the 15 realizations.
The mean and standard deviation of this average for each $\delta_L$ is collected and used to perform a $4^{\rm th}$-order (quartic) polynomial fit for the halo overdensity as a function of $\delta_L$. 
 The best fit values of the first- and second-order coefficients are then estimates of the linear and quadratic Lagrangian bias $b^L_1$ and $b^L_2$. The error on these estimates are obtained from the square root of diagonal elements of the covariance matrix recovered from the fit.

The corresponding Eulerian parameters $b_{n}$ can be obtained from the relation $(1+\delta_h^L)(1+\delta) = 1 + \Sigma_{n=1}^{\infty}(b_{n}/n!)\delta^{n}$ \citep{mw96} by substituting into it the approximate nonlinear $\delta(\delta_L)$ relation derived from spherical evolution: $\delta =\delta_{L}g(z)+(17/21)\delta_{L}^2g(z)^2+\mathcal{O}(\delta_L^3)$ \citep{Bernardeau92,wsck15b}, which leads to
\be
 \begin{split}
  b_{1} &= 1 + b_{1}^L g(z)^{-1},\\
  b_{2} &= b_{2}^L g(z)^{-2} + \frac{8}{21}b_{1}^L g(z)^{-1}\,.
 \end{split}
\ee
Here $g(z)\equiv D(z)/D(0)$ and $D(z)$ is the linear theory growth factor of the fiducial cosmology.

\subsection{Local cosmic web environment of haloes}
We use the {\it tidal anisotropy} variable $\alpha$ introduced by \citet{phs18a}
to quantify the halo's nonlinear local environment. We construct this from the eigenvalues $\lambda_1$, $\lambda_2$, $\lambda_3$ of the tidal tensor $\psi_{ij}\equiv {\partial^2\psi}/{\partial x^i \partial x^j}$, where $\psi$ satisfies the normalised Poisson equation $\nabla^2\psi = \delta$. The halo-centric $\alpha$ is then defined as
\be
 \alpha = \sqrt{q^2}/(1+\delta)\,,
 \label{eq:alpha-def}
\ee
where $q^2$ and $\delta$ are the halo-centric tidal shear 
\citep{hp88,ct96a} and overdensity respectively,
\begin{align}
 q^2 &= \frac{1}{2}[(\lambda_1-\lambda_2)^2+(\lambda_2-\lambda_3)^2+(\lambda_3-\lambda_1)^2] \,,\\
 \delta &= \lambda_1+\lambda_2+\lambda_3 \,.
\end{align}
The tidal anisotropy parameter $\alpha$ is, in general, a proxy for the anisotropy of the environment of a halo. Haloes with low $\alpha$ have highly isotropic local environments while those with high $\alpha$ reside in anisotropic filamentary environments. \citet{p20} provides theoretically motivated insights into the behaviour of $\alpha$.

\subsubsection{Measuring $\alpha$ in a fiducial $\delta_L=0$ simulation}
To compute $\alpha$ in the fiducial boxes, we start with the matter density field evaluated on an $844^3$ cubic lattice with comoving spacing $\simeq 0.51\,{\rm Mpc}$
using Cloud-in-Cell (CIC) interpolation. The density field is then Fourier transformed and Gaussian smoothed using a range  of smoothing scales $R_{G}$ to get the Fourier space field
${\delta}(\mathbf{k};R_{G}) = {\delta}({\mathbf{k}}) e^{-k^2R_G^2/2}$.
 Using this, we obtain the tidal tensor ${\psi}_{ij}(\mathbf{x};R_G)$ for various smoothing scales $R_{G}$ by inverting the normalised Poisson equation and taking derivatives,
 \be
 {\psi}_{ij}(\mathbf{x};R_G) = {\rm FT}\left[\frac{k_ik_j}{k^2}\,{\delta}(\mathbf{k};R_{G})\right].
 \ee
Then we compute the halo-centric tidal tensor by choosing, for each halo, the tidal tensor centred around the nearest lattice point $\mathbf{x}_{\rm halo}$ and then linearly interpolating between the two smoothing scales nearest to the scale $R_{\rm G,halo} = 4R_{200b}/\sqrt{5}$ \citep{phs18a}.
This scale has been chosen so as to have a larger $b_{1}\leftrightarrow\alpha$ correlation than the $b_{1}\leftrightarrow\delta$ correlation while minimising the $\alpha\leftrightarrow\delta$ correlation at fixed halo mass \citep[see Figure 5 of][]{phs18a}.

\subsubsection{Measuring $\alpha$ when $\delta_L\ne 0$}
In SU simulations where $\delta_L \ne 0$, we must account for certain subtleties while following our prescription for computing $\alpha$ as we discuss next. First is the issue of the units of length. The default unit of measuring length is comoving \Mpch. We would like to perform all computations in these units in the fiducial cosmology with $h=h_{\rm fid}$. However, at a cosmic time $t$ (redshift $z$ of the fiducial cosmology) our SU with $\delta_L\ne 0$ corresponds to a snapshot at redshift $\tilde z$ in an $N$-body simulation with a different Hubble constant $\tilde{h}$. To ensure that the \emph{proper} length of the smoothing scale is preserved across SU simulations, the units of length \emph{in the SU snapshot} are transformed as follows 
\be
\mathbf{x}\rightarrow \mathbf{x} \times \frac{h_{\rm fid}(1+z)}{\tilde{h}(1+{\tilde{z}})}\,.
\ee

\begin{figure*}
  \includegraphics[width=0.95\linewidth]{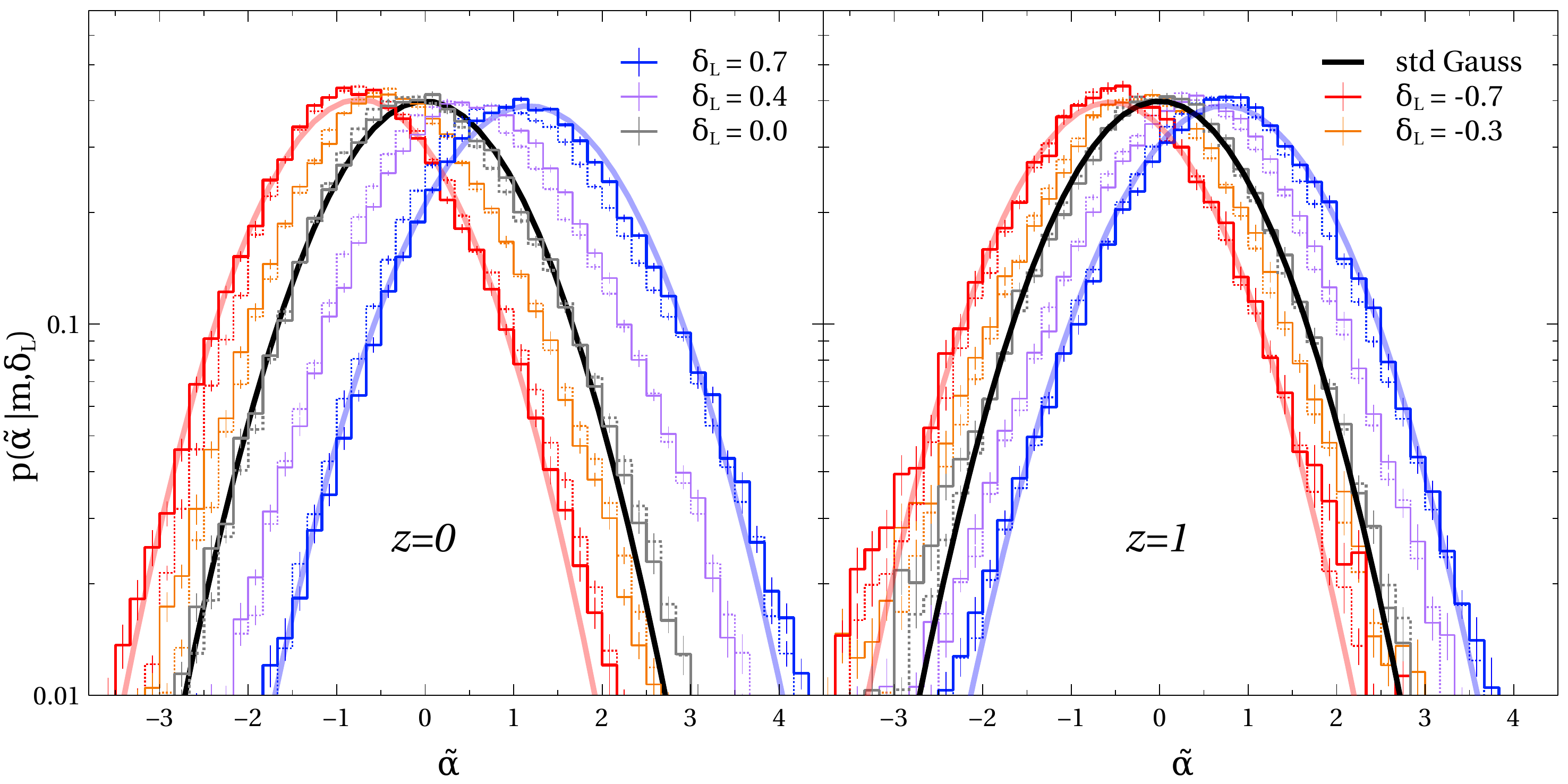}
 \caption{Probability distribution of tidal anisotropy $\tilde{\alpha}$ for different $\delta_L$, at z=0 ({\it left panel}) and z=1 ({\it right panel}), averaged over 15 realizations. Warmer (cooler) colors are used to denote $\delta_{L}<0\, (\delta_{L}>0)$ respectively and are detailed in the legend. The solid (dotted) line styles represent data from mass ranges $m = 6.2$-$10 \times 10^{12}\, (2$-$5 \times 10^{13})\,\Msun$. Solid black curve in each panel shows the standard Gaussian distribution $p(\tilde{\alpha}) = e^{-\tilde{\alpha}^2/2}/\sqrt{2 \pi}$ that we use to approximate the grey $\delta_{L}=0$ distribution in our analytical framework. The solid blue and red curves are other Gaussians with shifted mean and variance computed from direct measurements (see equations~\ref{eq:mu} and \ref{eq:sigma}), also used to approximate distribution of $\tilde{\alpha}$ in $\delta_L=0.7$ and $\delta_L=-0.7$ respectively.}
 \label{distribution-alpha}

\end{figure*}

\noindent
Secondly, we need to modify our CIC algorithm for computing the overdensity field $\delta(\mathbf{x})$. Recall that the overdensity can be written as,
\be
 \delta_{x_n} = \dfrac{\varrho - \bar{\varrho}}{\bar{\varrho}}= \frac{\Delta n_{x_{n}}}{N_{\rm part}} N_{\rm Grid} - 1.
 \label{cic_old}
\ee
where $\Delta n_{x_n}$ is the number of dark matter particles contributing to the lattice point $x_n$, $N_{\rm part}$ is the total number of dark matter particles and $N_{\rm Grid}$ is the total number of lattice points. Recollect that in order to go from the first equality to the next in \eqn{cic_old}, we assume that the average density $\varrho_{\rm sim}$ of the simulation box is equal to the average density $\bar{\varrho}$ of the fiducial universe. However, this is the case only for simulations where $\delta_L=0$. In the simulations with positive $\delta_L$, $\varrho_{\rm sim}>\bar{\varrho}$ and when $\delta_L$ is negative, $\varrho_{\rm sim}<\bar{\varrho}$ . The CIC overdensity after accounting for this can be computed as
\be
  \delta_{x_n} =  \frac{\Delta n_{x_{n}}}{N_{\rm part}} \dfrac{(1+\tilde{z})^3}{(1+z)^3}N_{\rm Grid} - 1\,.
 \label{cic_su}
\ee
Lastly, since different SU boxes have different lengths in our default units, we alter $N_{\rm Grid}$ so as to keep the grid size equal. This tuning of $N_{\rm Grid}$ will keep the CIC density field calculation consistent across different SU simulation boxes. We had taken $N_{\rm Grid}=844^3$ for $\delta_L=0$ simulations, and for other simulations, we alter $N_{\rm Grid}$ to be $844^3(1+z)^3/(1+\tilde{z})^3$ rounded to the nearest integer\footnote{The first two effects are relatively important while the last effect is of lesser importance. This is because the first two modifications lie at the centre of the SU approach while the last effect plays a significant role only if $\alpha$ has not converged.}. 

Following the prescription above, we compute $\alpha$. We also perform convergence tests to ensure that our measurement values have sufficiently converged.

\subsubsection{Distribution of $\alpha$}

\label{sec:alpha-distribution}

\begin{figure*}
\includegraphics[width=0.95\linewidth]{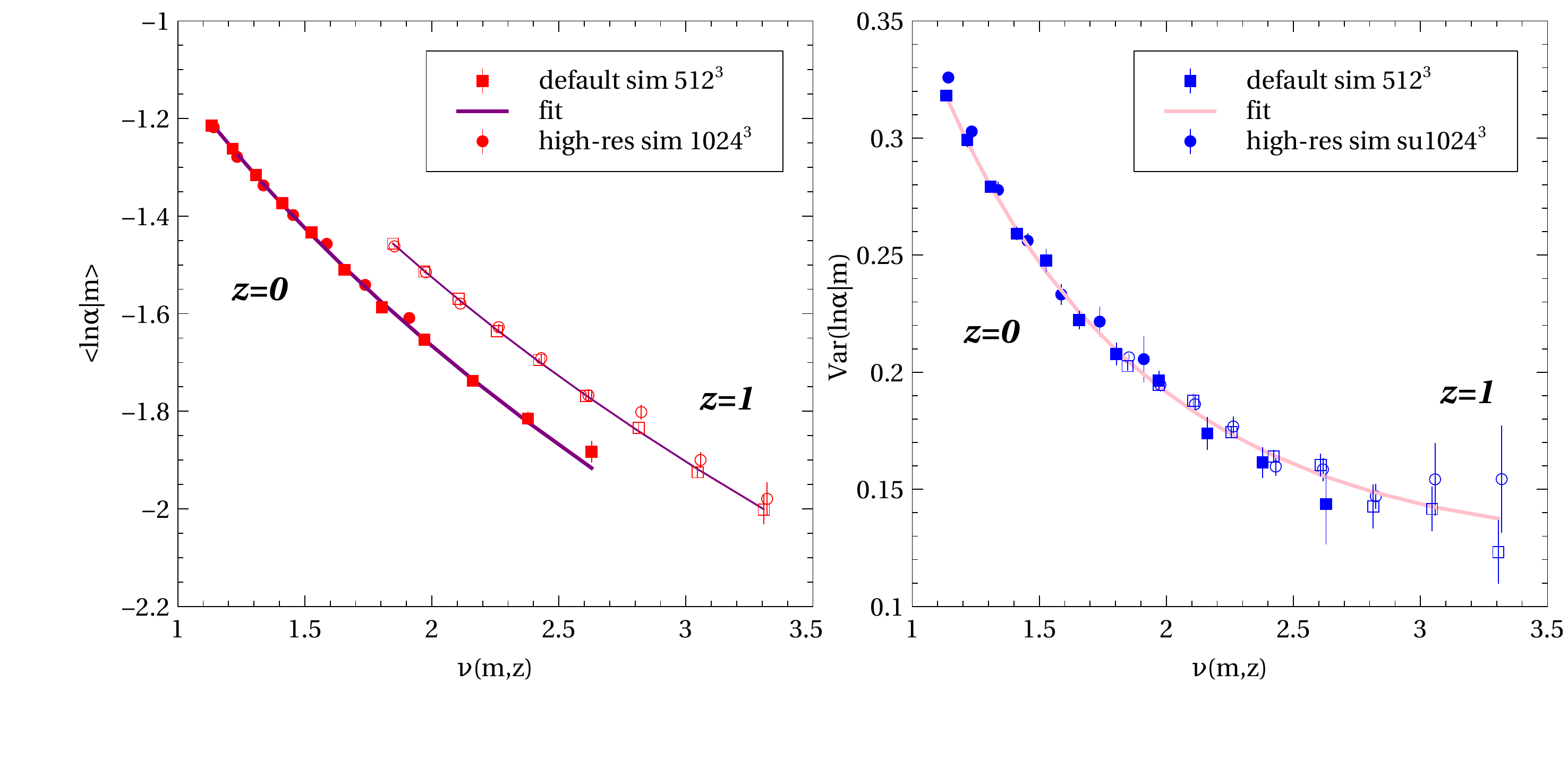}
\caption{Mean and variance of tidal anisotropy $\alpha$ in the fiducial cosmology as a function of peak height $\nu$ for redshifts z=(0)1 as indicated by filled(empty) markers. The data with square markers is computed in the default simulation box having particle count $512^3$  while the data with circular markers is computed in a higher resolution box having particle count $1024^3$. The solid curves are obtained by fitting polynomials as a function of $y=\log_{10}(\nu/2.05)$ to the mean (variance) in the \emph{left (right)} panel, as described in section~\ref{sec:alpha-distribution}.
 Best fit values and errors are given in Table~\ref{table:alpha_mean_variance}. }
\label{fig:alphaVmass}
\end{figure*}

\begin{table}
\centering
\begin{tabular}{cccccc}
 \hline 
 \hline  
&&&&&$\chi^2$\\ 
&$S_{0}$&$S_{1}$&$S_{2}$&& (17\,d.o.f.)\\ 
\hline 
value&0.187&-0.359&0.572&&10.44 \\ \hline 
std dev&0.001&0.012&0.058& \\ \hline 
corr $S_{0}$&1.000&-0.060&-0.366& \\ 
corr $S_{1}$&-&1.000&0.874& \\ 
 \hline\hline \\ 
&&&&& $\chi^2$ \\ 
&$m_{00}$&$m_{10}$&$m_{1}$&$m_{2}$& (16\,d.o.f.) \\ 
\hline 
value&-1.688&-1.547&-2.038&-0.706& 14.72 \\ \hline 
std dev&0.003&0.001&0.020&0.084& \\ \hline 
corr $m_{00}$&1.000&-0.018&0.502&-0.187& \\ corr $m_{10}$&-&1.000&-0.210&-0.231& \\  
corr $m_1$&-&-&1.000&0.684& \\ 
\hline\hline \\ 
\end{tabular}
\caption{Best fit coefficients and covariance matrices of quadratic polynomial fits $\mu_{0}$ and $\sigma_{0}$ as a function of logarithmic peak height $y= \log_{10}(\nu/2.05) $. Fits were performed with the coefficients defining $\sigma_{0}^2 = S_0+S_1y+S_2y^2$ and a 4 parameter joint fit for both redshift $z=0$ and $z=1$ as follows, $\mu_{0}(z=0) = m_{00}+m_1y+m_2y^2$ and $\mu_{0}(z=1) = m_{10}+m_1y+m_2y^2$. Upper and lower blocks correspond to fits for $\mu_0$ and $\sigma_0$ respectively. In each block the first row gives the best fit values, the second row gives the standard deviation and the last few rows give the correlation coefficients.}
\label{table:alpha_mean_variance}
\end{table}

\noindent
The tidal anisotropy $\alpha$, for populations in narrow mass ranges, can be Gaussianized by a relatively simple transformation as it has a near-Lognormal distribution. For each mass bin, we can standardize the tidal anisotropy $\alpha$ as follows,
\be      
 \tilde{\alpha} \equiv \dfrac{\rm \ln \alpha-\mu_{0}}{\sigma_0},
\label{eq:redefinealpha}
 \ee
where
\begin{align}
 \mu_{0} &\equiv \avg{\ln \alpha|m,\delta_L =0} \,,
\label{eq:mu0}\\
\sigma_0^2 &\equiv {\rm Var} (\ln\alpha|m,\delta_L=0) \,.
\label{eq:sigma0}
\end{align}
Thus, if the distrbution of $\alpha$ were exactly Lognormal, $\tilde{\alpha}$ by construction would have a standard Gaussian distribution in the $\delta_L=0$ universe.\footnote{Note that $\sigma_0$ defined here should not be confused with the standard deviation of linear density fluctuation.} This can in fact be seen in Figure~\ref{distribution-alpha}, where the grey histogram showing the distribution of $\tilde{\alpha}$ in the $\delta_L=0$ universe is well approximated by the thick solid black standard Gaussian. However, from the blue and red step histograms of the same figure, we see that this is not the case for  $\delta_L\ne0$ universe. From experimenting with the simulation data for $\delta_L\ne0$, we find that $
\tilde{\alpha}$ as defined above is still approximately Gaussian distributed but with a systematic shift in mean and variance as $\delta_{L}$ becomes progressively positive or negative. This observation encourages us to define the mean and variance for a mass range as a Taylor expansion in powers of $\delta_L$ \citep{pp17}, 
\begin{align}
 \mu(m,\delta_L)&\equiv \sum_{n=1}^{\infty}\frac{\mu_{n}^{L}(m)}{n!}\delta_L^n\,,
\label{eq:mu}\\
 \sigma^2(m,\delta_L)&\equiv 1 + \sum_{n=1}^{\infty}\dfrac{\Sigma_n^L(m)}{n!}\delta_L^n\,.
\label{eq:sigma}
\end{align}
Figure~\ref{fig:alphaVmass} shows equations~\eqref{eq:mu0} and \eqref{eq:sigma0} as a function of `peak height' $\nu(m,z)$ for $\delta_{L}=0$\footnote{The peak height is defined as $\nu(m,z)\equiv \delta_c(z)/\sigma_{0}(m)$, where $\delta_c(z)$ is the critical threshold for spherical collapse and $\sigma_0(m)$ is the standard deviation of linear fluctuations smoothed on Lagrangian radius scale, both linearly extrapolated to $z=0$ (so $\delta_c(z)=1.686/g(z)$).}. In the right panel, the data describing redshift 0 and 1 are combined and fit with a universal quadratic polynomial describing the variance of logarithmic tidal anisotropy using $\sigma_0^2=S_0 + S_1 y + S_2 y^2$. Here $y=\log_{10}(\nu/2.05)$ is the logarithmic peak height. In the left panel, a 4-parameter joint fit is performed on the mean value of tidal anisotropy to the polynomial $\mu_{0}(y,z) = m_{00}(1-z)+m_{10}z +m_1 y + m_2 y^2 $. Thus we have two polynomials corresponding to two data sets at redshift 0 and 1 respectively. The joint fit is produced by minimising the sum of the individual chi-squared functions. Table~\ref{table:alpha_mean_variance} provides the best fit values and covariance matrix for these fits.

The discussions in this section will be useful in  subsequent sections where we discuss an analytical framework relying on a model for the distribution of $\tilde{\alpha}$. 

\section{Framework for high-precision bias calibration}
\label{sec:alpha_calibration}
\subsection{Lognormal Model}
\label{sec:lnmodel}
This section is a straightforward utilization of the analytic framework developed by \citet{pp17}, which we will refer to as the Lognormal model for halo assembly bias. Here we use the tidal anisotropy $\tilde{\alpha}$ from \eqn{eq:redefinealpha} 
as the assembly bias variable. We can include the dependence of the bias coefficients on $\tilde{\alpha}$ in \eqns{eq:del-num} and \eqref{eq:del-delL} and write as 

\be
\begin{split}
 \delta_{h}^{L}(m,\tilde{\alpha},\delta_L)&\equiv \dfrac{n(m,\tilde{\alpha}|\delta_L)}{n(m,\tilde{\alpha}|\delta_{L}=0)}-1\,.\\
  &=\sum_{n=1}^{\infty}\dfrac{b_n^L(m,\tilde{\alpha})}{n!}\delta^n_L\,.
  \end{split}
 \label{eq:del-num-alpha}
\ee
Combining equations \eqref{eq:del-num}, \eqref{eq:del-delL} and \eqref{eq:del-num-alpha}, we can write the dependence of bias coefficients on $\tilde{\alpha}$ in terms of its probability distribution $p(\tilde{\alpha}|m,\delta_L)$
\be
\begin{split}
1+\sum_{n=1}^{\infty}\dfrac{b_n^L(m,\tilde{\alpha})}{n!}\delta^n_L &= \left(1+\sum_{n=1}^{\infty}\dfrac{b_n^L(m)}{n!}\delta^n_L\right)
\dfrac{p(\tilde{\alpha}|m,\delta_L)}{p(\tilde{\alpha}|m,\delta_L=0)}.
\end{split}
\ee
In the above, we have used Bayes' theorem to express the number density of haloes in terms of the distribution of $\tilde{\alpha}$ as $n(m,\tilde{\alpha}|\delta_L) = n(m|\delta_L) p(\tilde{\alpha}|m,\delta_L)$.
As discussed in Section~\ref{sec:alpha-distribution}, the probability distribution of $\tilde{\alpha}$ for a fixed mass $m$ and $\delta_L$ is a Gaussian with mean $\mu$ and variance $\sigma$ and can be expressed in powers of 
$\delta_L$ as shown in \eqns{eq:mu} and \eqref{eq:sigma}. Hence it is possible to write out the above expression in powers of $\delta_L$ and equate the coefficients of each power to obtain equations for the dependence of each bias coefficient on $\tilde{\alpha}$.
In particular, the Lagrangian linear and quadratic bias can be expressed as 
\be
 b_{1}^L(m,\tilde{\alpha}) = b_{1}^{L}(m) + \mu_{1}^{L}(m) H_{1}(\tilde{\alpha}) + \dfrac{1}{2}\Sigma_{1}^{L}(m) H_{2}(\tilde{\alpha}),
 \label{eq:bias(m,s)}
\ee
\be
\begin{split}
b_{2}^{L}(m,\tilde{\alpha}) &= b_{2}^{L}(m) +\{\mu_{2}^L(m)+2b_{1}^L(m)\mu_{1}^L(m)\}H_{1}(\tilde{\alpha})\\
&+\{\mu_{1}^L(m)^2 + b_{1}^L(m)\Sigma_{1}^L(m)+\frac{1}{2}\Sigma_{2}^L(m)\}H_{2}(\tilde{\alpha})\\
&+\mu_{1}^L(m) \Sigma_{1}^L(m)  H_{3}(\tilde{\alpha})+\frac{1}{4}\Sigma_{1}^L(m)^2 H_{4}(\tilde{\alpha}). 
\end{split}
\label{eq:bias2(m,s)}
\ee
where $H_{n}$ are the `probabilist's' Hermite polynomials (equation \ref{eq:hermite}), and $\mu_{n}^{L}(m)$ and $\Sigma_{n}^{L}(m)$ are coefficients as they occur in \eqns{eq:mu} and \eqref{eq:sigma} \citep[see Appendix C of][for a derivation]{pp17}. 
In Section~\ref{sec:musigcoeff}, we describe how to obtain these coefficients from simulations as continuous functions of mass and redshift. Once we fit the $\tilde{\alpha}$-independent $b^{L}_{1}(m)$ and $b^{L}_{2}(m)$ exactly as described in \citet{lwbs16}, equations~\eqref{eq:bias(m,s)} and \eqref{eq:bias2(m,s)} enable us to provide a continuous prediction for the dependence of bias on both mass and tidal anisotropy. 

\subsection{Obtaining Taylor Coefficients of \texorpdfstring{$\mu$ and $\sigma$}{Mu and sigma}}
\label{sec:musigcoeff}

\begin{figure*}
\centering
 \includegraphics[width=0.95\textwidth]{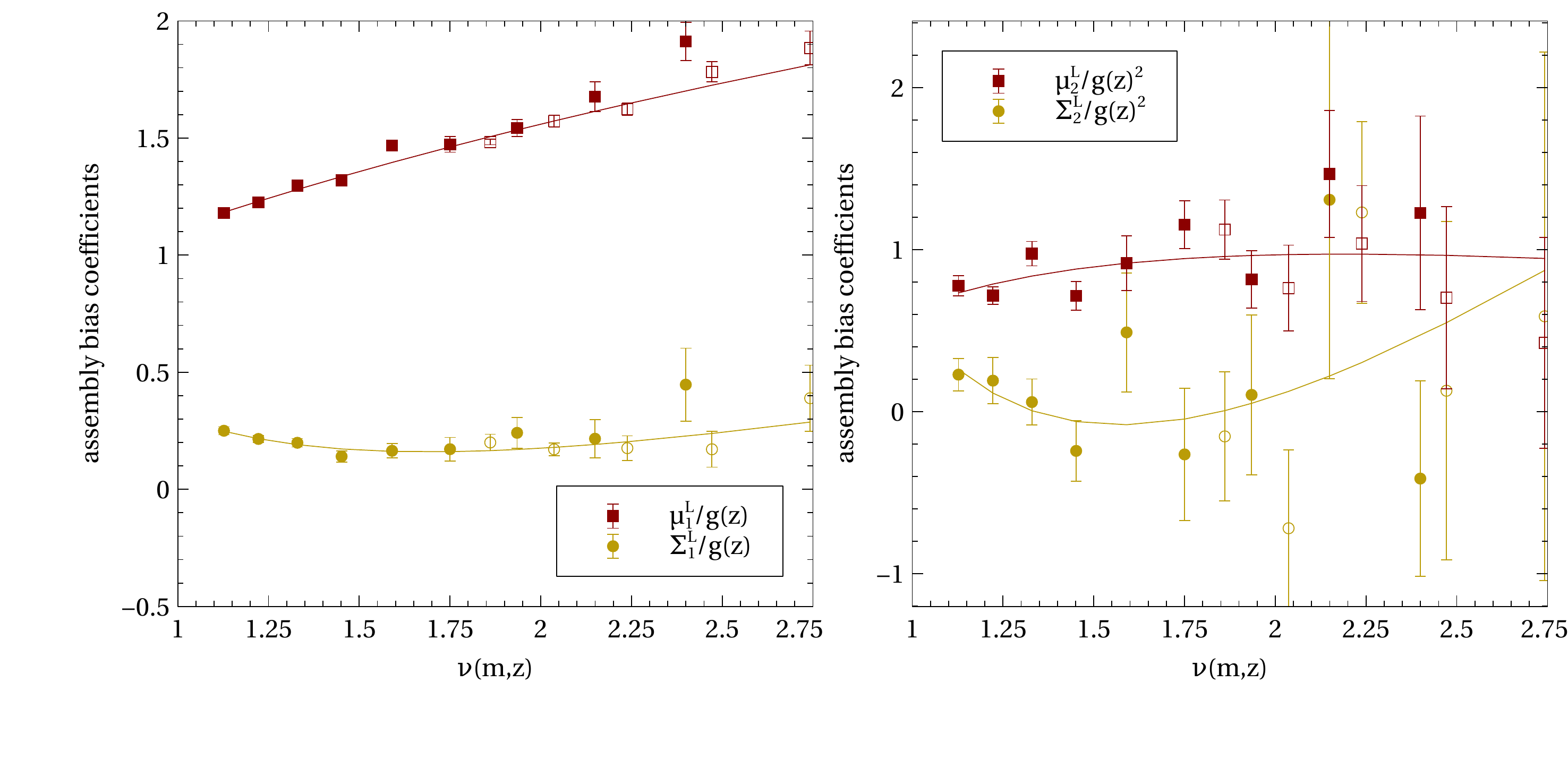}
 \caption{Lagrangian assembly bias coefficients $\mu_n^L$ and $\Sigma_n^L$ (equations \ref{eq:mu} and \ref{eq:sigma}), extrapolated to the measurement redshift by dividing by $g(z)^n$ and shown as functions of $\nu(m_{200b},z)$, for $n=1$ \emph{(left panel)} and $n=2$ \emph{(right panel)}.
The points with error bars show measurements from simulations (details in section~\ref{sec:computeb1}). 
 The filled (empty) symbols show measurements at $z=0$ ($z=1$).
 The solid curves are obtained by fitting a quadratic polynomial as a function of $y = \log_{10}(\nu/1.5)$ using the points and errors in the range $1.1<\nu<2.8$. The best fit values and errors from this quadratic fit are given in Table~\ref{table:alpha}.
}
 \label{alphamusigma}
\end{figure*}

\noindent
In the simulations, we compute $\mu(m,\delta_L)$ and $\sigma(m,\delta_L)$
for each $\delta_L$ and perform a least-squares fit on \eqns{eq:mu} and \eqref{eq:sigma} truncated at $4^{\rm th}$ order in $\delta_L$ (i.e., quartic polynomial fits), as discussed next.

For each of the 15 realizations, we take the halo population corresponding to an overdensity $\delta_L$ 
and mass bin $m$ and compute mean and central 68.3\% scatter of $\tilde{\alpha}$. We estimate errors on these quantities using 50 bootstrap resampled populations. The blue and red smooth curves in Figure~\ref{distribution-alpha} show how Gaussians with mean $\mu(m,\delta_L)$ and variance $\sigma^2(m,\delta_L)$ compare with the actual distribution of $\tilde{\alpha}$ in the simulation. For each realization, we fit a $4^{\rm th}$ order polynomial for the dependence of  $\mu(m,\delta_L)$ and $\sigma^2(m,\delta_L)$ on $\delta_L$ using the errors calculated in the previous step, and retain the coefficients corresponding to $\delta_L$ and $\delta_L^2$.  Thus, we have one set of fitting coefficients $\mu_{1}^L(m)$, $\mu_{2}^L(m)$, $\Sigma_1^L(m)$ and $\Sigma_2^L(m)$ (see equations~\ref{eq:mu} and~\ref{eq:sigma}) for each  of the 15 realizations. 

It is convenient to combine the dependence on mass and redshift in these coefficients into a single dependence on peak height $\nu(m,z)$. This unification is done by noting that defining $\mu_{n}^L(m,z) \equiv \mu^{L}_{n}(m)g(z)^{-n}$
and $\Sigma_{n}^L(m,z) \equiv  \Sigma_{n}^L(m)g(z)^{-n}$ makes the coefficients universal functions of $\nu$, as shown in
Figure~\ref{alphamusigma} where the points show the mean over 15 realizations of these coefficients as a function of peak height.
The error bars show the standard error over the mean.
We further fit these points by quadratic polynomials in $\log_{10}(\nu)$, shown as the solid curves in the Figure.
The degree of the polynomial is chosen after analysis with the AIC criterion \citep{akaike1974,sugiura1978}. While the fits on $\mu_{1}^{L}$,$\mu_{2}^{L}$,$\Sigma_{1}^L$ are reasonable, we note that the scatter in $\Sigma_{2}^L$ is larger than the errorbars especially at higher masses where the number of haloes are smaller, possibly because of the probability distribution function of $\alpha$ not having converged. We are also ignoring the covariances between the coefficients $\Sigma_{1}^{L},\Sigma_{2}^{L}$ that could potentially affect the errorbars. Table~\ref{table:alpha} gives the resulting fitting coefficients and covariance matrices. This table is useful in computing error bars for the Lognormal model as can be seen in the next section.

\begin{table*}
\centering
\begin{tabular}{@{}llllllllllll@{}}
 \hline 
 \hline \\ 
&$\mu_{10}$&$\mu_{11}$&$\mu_{12}$&$\chi^2(\rm 10\,d.o.f.)$& & & $\Sigma_{10}$ & $\Sigma_{11}$ & $\Sigma_{12}$ & $\chi^2(\rm 10\,d.o.f.)$ &  \\ 
 \hline
value     &  1.357 & 1.507 & 0.899 & 20.025    &  &       & 0.168 &-0.304  &2.902 &  5.660 \\ 
std dev     &  0.008 & 0.049 & 0.555 &     &  &       & 0.013 &0.089  &0.969 &   \\ 

 \hline 
corr $\mu_{10}$&  1.0 & 0.368 & -0.784 &    &  &   corr $\Sigma_{10}$    & 1.0 &0.457  &-0.773 &   \\ 
corr $\mu_{11}$&  - & 1.0& -0.258 &   &  &  corr $\Sigma_{11}$      & - & 1.0  &-0.319 &   \\ 
\hline 
 \hline \\ 
 & $\mu_{20}$ & $\mu_{21}$ & $\mu_{22}$ &  $\chi^2(\rm 10\,d.o.f.)$ &  &              & $\Sigma_{20}$ & $\Sigma_{21}$ & $\Sigma_{22}$ &  $\chi^2(\rm 10\,d.o.f.)$ &  \\ 

 \hline 
value     &  0.889 & 0.968 & -2.365 & 15.843    &  &       & -0.077 &-0.745  &15.833 &  10.422 \\ 
std dev     &  0.053 & 0.412 & 4.529 &     &  &       & 0.111 &0.843  &8.634 &   \\ 

 \hline
corr $\mu_{20}$&  1.0 & 0.389 & -0.702 &    &  &   corr $\Sigma_{20}$    & 1.0 &0.384  &-0.675 &  \\ 
corr $\mu_{21}$&  - & 1.0 & 0.022 &   &  &  corr $\Sigma_{21}$      & - &1.0  &0.144 &  \\ 
\hline 
 \hline \\ 
\end{tabular}
\caption{ Best fit coefficients and covariance matrices of quadratic polynomial fits to $\mu^{L}_{n}(y)$ and $\Sigma^{L}_{n}(y)$ as a function of logarithmic peak height $y \equiv \log_{10}[\nu(m,z)/1.5]$ for n=1,2 (See Figure~\ref{alphamusigma}). The fits were performed in the range 1.1$\le\nu \le 2.8$ with the coefficients defining $\mu_n^L/g = \mu_{n0} + \mu_{n1}y + \mu_{n2}y^{2} $ and $\Sigma_n^L/g^2 = \Sigma_{n0} + \Sigma_{n1}y + \Sigma_{n2}y^{2} $. The upper and lower blocks give these polynomial coefficients for $n=1,2$  respectively. In each block, the first row gives the least squares best fit values, the second row gives the standard deviation (square root of the diagonal elements of the covariance matrix). The last two rows give the correlation coefficients (elements of the covariance matrix $C_{ij}$ divided by $\sqrt{C_{ii}C_{jj}}$).}
\label{table:alpha}
\end{table*}

\subsection{Linear halo bias and tidal anisotropy}

\begin{figure}
\center
 \includegraphics[width=\linewidth]{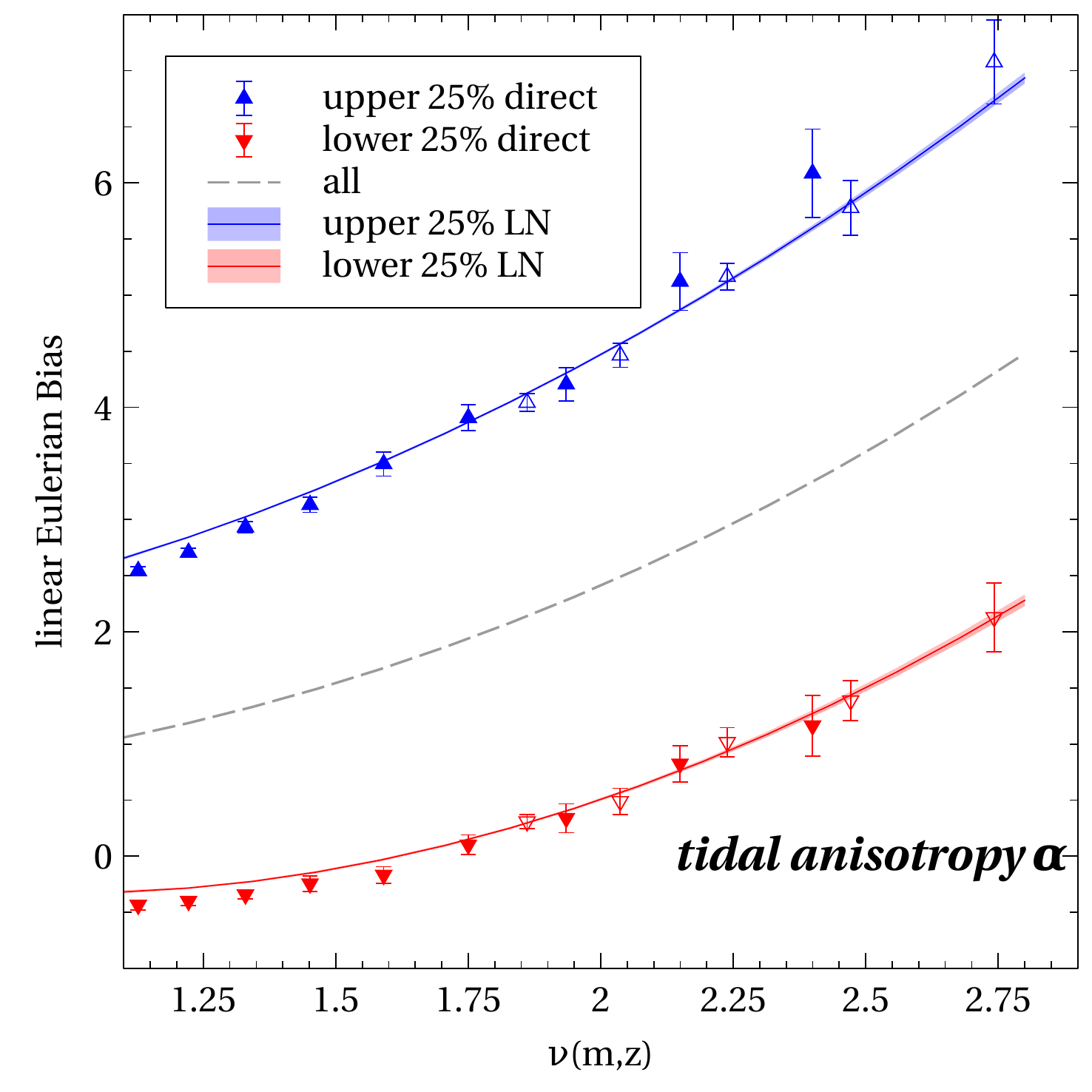}
  \caption{Linear halo bias $b_1$ as a function of peak height $\nu$ for upper and lower quartiles of $\alpha$ (i.e., $\tilde{\alpha}>0.675$ and $\tilde{\alpha}<-0.675$, see equation~\ref{eq:redefinealpha}). The data points with error bars are obtained from simulations (see section~\ref{sec:computeb1}). The two solid curves are obtained by taking the Lognormal model $b_1^L(m,\tilde{\alpha})$ and averaging within the upper and lower quartile of $\tilde{\alpha}$. The covariance matrix from Table~\ref{table:alpha} is used to sample $\mu_{1}^L$ and $\Sigma_1^L $ 300 times and the standard deviation of $b_1^L(m,\tilde{\alpha})$ computed from each of these times is plotted as an error band around the solid curves. The black dashed curve shows the analytic fit for the linear bias of all haloes from \citet{Tinker10}.}
 \label{fig:ab_alpha}
\end{figure}

\noindent
We now compare with known results for the dependence of linear halo bias on $\alpha$.
One conventional way in which assembly bias is visualised is to compute the mean halo bias in the upper and lower quartiles of the assembly bias variable for each mass bin. 
Since $\tilde{\alpha}$ follows a standard Gaussian distribution (see Figure~\ref{distribution-alpha}), these quartiles correspond to halo populations with $\tilde{\alpha}>0.675$ and $\tilde{\alpha}<-0.675$.

The solid curves in Figure~\ref{fig:ab_alpha} show the Lognormal model for $b_1$ applied to these two populations; these are obtained using \eqn{eq:bias(m,s)} averaged over the quartiles of $\tilde{\alpha}$ weighted by the standard Gaussian distribution. These are used along with best fit values of the coefficients $\mu_{1}^L$ and $\Sigma_1^L $ from Table~\ref{table:alpha}. 
The error covariance of these coefficients is used to generate an error band around the solid curves by Monte Carlo sampling the coefficients and computing the standard deviation of the resulting $b_1$.

\begin{figure}
\center
 \includegraphics[width=\linewidth]{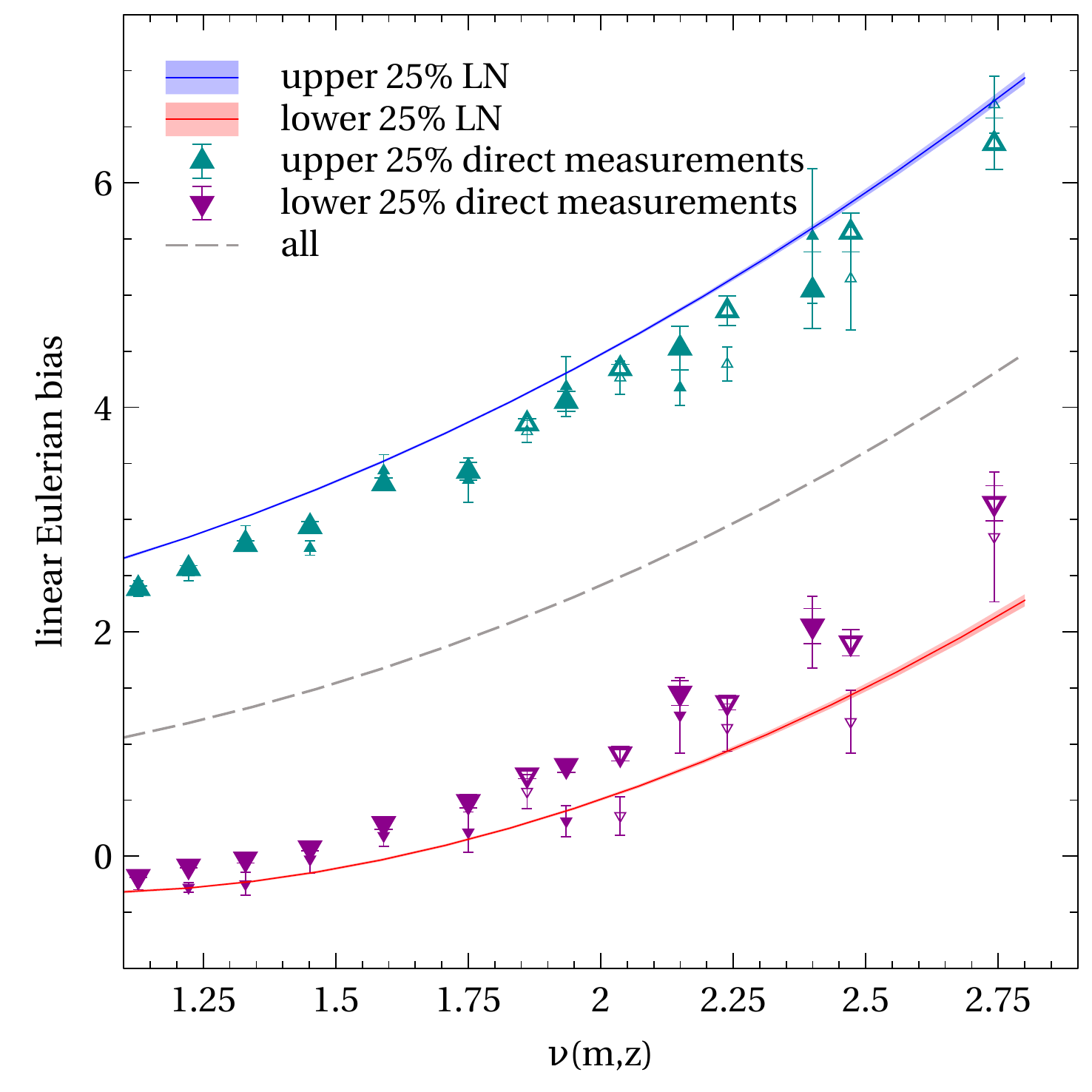}
 \caption{Linear halo bias $b_1$ as a function of peak height $\nu$ for upper and lower quartiles of $\alpha$ population. The data points with error bars are obtained from direct measurement of halo bias in the simulations (essentially, a weighted mean of low-$k$ measurements of the ratio of the halo-matter cross power spectrum to the matter auto-power spectrum). The thicker (thinner) markers show linear bias measurements with $k/\Mpcinvh<0.1(0.03)$.  The solid curves with error bands around it are the same as in Figure~\ref{fig:ab_alpha}.}
 \label{fig:ab_directmmn}
 \end{figure}

For comparison, we also compute the peak-background split bias described in Section~\ref{sec:computeb1} for the halo populations with $\tilde{\alpha}>0.675$ and $\tilde{\alpha}<-0.675$ separately. The results, shown as the two sets of points with error bars in  Figure~\ref{fig:ab_alpha}, agree well with the Lognormal model, but with larger errors. Thus, the Lognormal model is a very convenient noise reduction technique for computing halo assembly bias, as noted previously by \citet{pp17}.
In Figure~\ref{fig:ab_directmmn} we compare the Lognormal model to direct computation of linear halo bias using low-$k$ ($0.02\le k/(\Mpcinvh) <0.1$) measurements of the ratio of halo-matter cross power spectrum to the matter auto-power spectrum. We see that the direct measurements broadly agree with the SU results showing the same qualitative trends with  overall reduced strength. The quantitative differences between the two are likely due to the fact that the SU approach probes the infinite wavelength $k\to0$ modes while any direct measurement will be limited by the size of the simulation box considered. The halo bias is also computed in a smaller range of $k$ modes ($0.02\le k/(\Mpcinvh) <0.03$) and shown in the same figure with thinner markers. While sample variance makes these measurements noisier, the agreement with the SU result improves, thus demonstrating the susceptibility of direct halo bias measurements to the scale dependence of bias. 

\begin{figure}
    \centering
    \includegraphics[width=\linewidth]{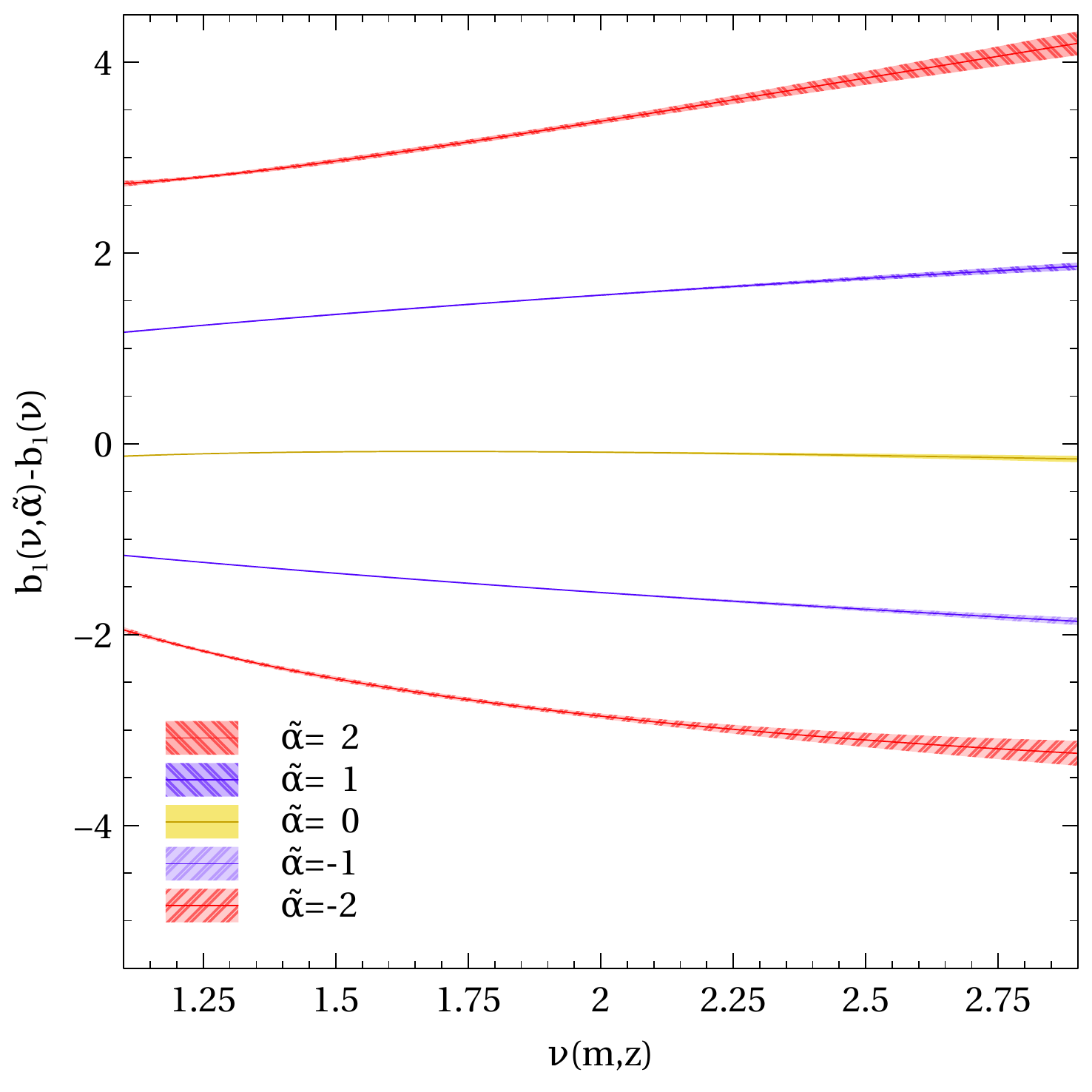}
    \caption{Assembly bias at fixed standardised tidal anisotropy $\tilde{\alpha}$. Each curve is obtained from the Lognormal model by taking the difference $b_1(\nu,\tilde{\alpha})-b_1(\nu)$ (see equation~\ref{eq:bias(m,s)}) at $\tilde{\alpha}=\pm2$ (red), $\tilde{\alpha}=\pm 1$ (purple) and $\tilde{\alpha}=0$ (yellow). The error bands are computed with the same procedure as described in Figure~\ref{fig:ab_alpha}.}
    \label{fig:b1fixedalpha}
\end{figure}

We emphasize that the analysis in this section, though interesting for comparing with literature, does not demonstrate the full capability of the Lognormal formalism. The formalism allows for the calculation of bias at fixed values of $\alpha$ and $\nu$, which is much more informative than binning in arbitrary percentiles. This feature has been shown in Figure~\ref{fig:b1fixedalpha}  as the difference between $b_{1}(\nu,\tilde{\alpha})-b_{1}(\nu)$ for a few fixed values of $\alpha$. For example, the curve labelled $\tilde{\alpha}=0$ represents how much the linear bias of the population of halos in the $50^{\rm th}$ percentile of $\alpha$ distribution differs from the mean bias of the whole population in every mass range. Though this curve is close to zero it should be noted that, in general, there is no reason why setting $\tilde{\alpha}=0$ should be equivalent to taking an average over the entire distribution of $\tilde{\alpha}$. This is simply a feature of the non-linear, monotonic relation between bias and $\tilde{\alpha}$\footnote{This should also be clear from examining the analytical expression in equation~\ref{eq:bias(m,s)}, for example, which explicitly depends on $\tilde{\alpha}^2$ through $H_{2}(\tilde{\alpha})$. Averaging this and setting $\tilde{\alpha}=0$ are not equivalent, since $\avg{ H_{2}(\tilde{\alpha})}=0$ while $H_{2}(\tilde{\alpha}=0)=-1$}. It is also interesting to note here that the strength of assembly bias in Figure~\ref{fig:b1fixedalpha} is almost a constant with peak height for lower $|\tilde\alpha|$ values, which emphasizes the point made by \citet{phs18a} that tidal anisotropy appears to be more relevant in determining linear halo bias than is halo mass.


\begin{figure}
\center
 \includegraphics[width=\linewidth]{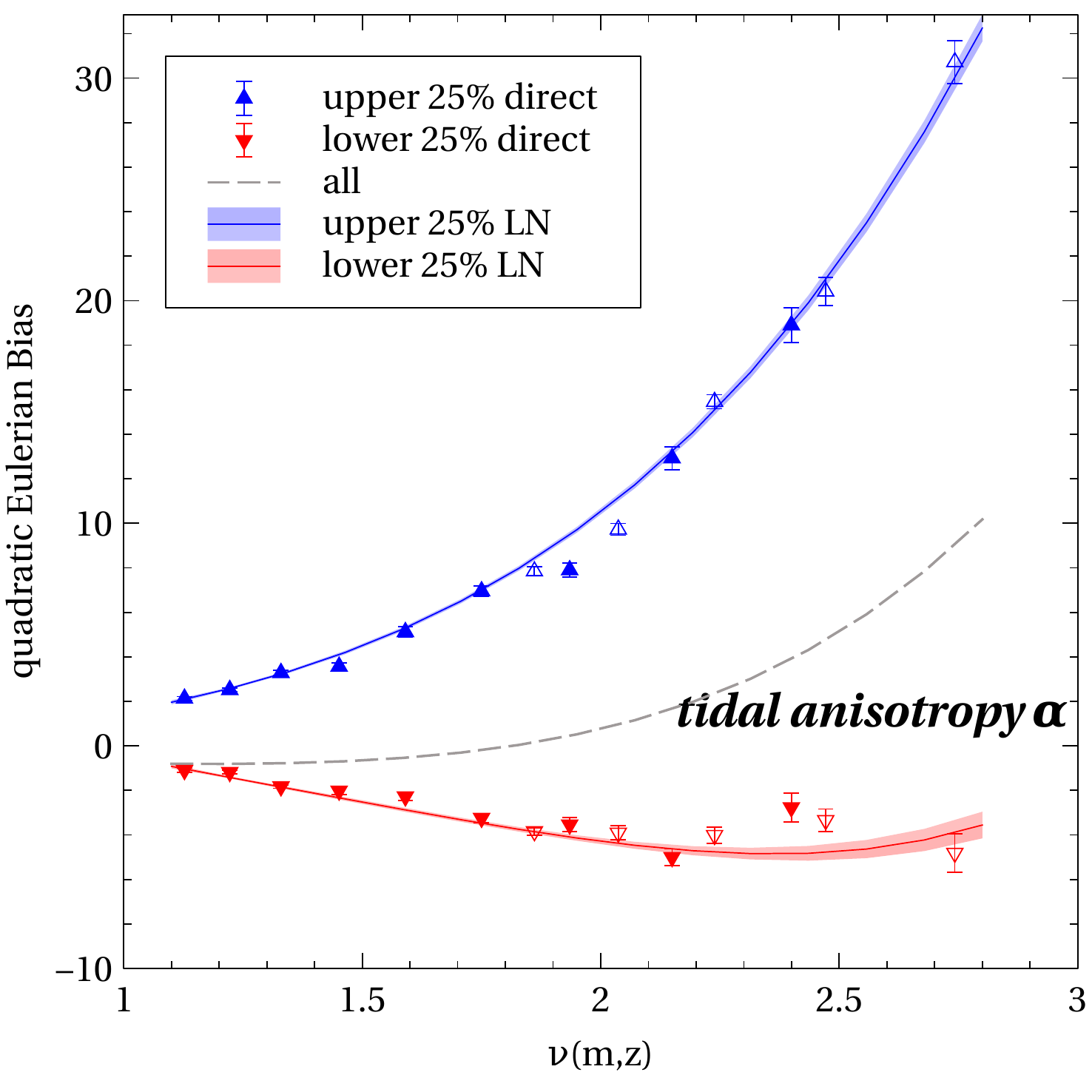}
  \caption{Quadratic halo bias $b_2$ as a function of peak height $\nu$ for upper and lower quartiles of $\alpha$.
  The points and curves are formatted identically to Figure~\ref{fig:ab_alpha} and show results from the simulations and the Lognormal model, respectively, for $b_{2}(m,\tilde{\alpha})$.
  The dashed black line shows the fitting function for the full halo population from equation 5.2 of \citet{lwbs16}. See text for a discussion.}
 \label{fig:ab2_alpha}
\end{figure}

\subsection{Quadratic halo bias and tidal anisotropy}
The quadratic assembly bias with respect to parameters beyond halo mass has been studied previously for halo properties like concentration, spin, mass accretion rate, and ellipticity (\citealp{abl08,lms17}). The dependence of quadratic bias $b_{2}$ on tidal anisotropy is expected on general grounds but has not, to our knowledge, been demonstrated before. We do so in this section; both the measurements and the analytical framework above are set up to effortlessly obtain the quadratic bias in addition to the linear bias.

Figure~\ref{fig:ab2_alpha} shows the difference in $b_2$ for the upper and lower quartiles of the tidal anisotropy $\alpha$. Interestingly, the upper and lower quartiles have opposite signs in all the mass ranges. The upper quartile population having positive values is expected from the extreme non-Gaussianities and non-linearities present in the spatial distribution of haloes in dense filamentary (high $\tilde{\alpha}$) environment. The near-zero, slightly negative $b_2$ of haloes in isotropic regions (low $\tilde{\alpha}$ quartile) is more complicated, as it could either have negative skew from being in an underdense void or a positive skew from being in an overdense cluster. There are many examples of tracers that have negative $b_2$ (\citealp{fffs2001}; \citealp{gj09}; \citealp{hmxmw19}). We can see that the dependence on the environment is clearly strong; the relative difference between any quartile $b_{1}$ and mean $b_1$ is of the order of unity while the relative difference between any quartile and mean $b_2$ is of the order of 10. 

Unlike $b_1$, the $\alpha$-dependence of $b_2$ is also a strong function of $\nu$, consistent with the expectation that $b_2$ depends on significantly more nonlinear scales than does $b_1$. It should also be clear from \eqn{eq:bias2(m,s)} that, similar to $b_1$, our formalism allows for the computation of $b_2$ at fixed $\tilde\alpha$, not just in bins of $\tilde\alpha$. We do not show these results for brevity.

\section{Extension to other secondary properties}
\label{sec:conc_ab}
Previously, \citet{rphs19} have considered direct halo-by-halo measurements of linear bias $b_{1}$ in standard $N$-body simulations \citep{phs18a} and internal property $c$ as random variables, allowing the correlation between them at fixed halo mass to be defined as assembly bias. \citet{rphs19} showed that the halo bias and internal property are consistent with being conditionally independent given the tidal anisotropy,\footnote{The previous result was with $\alpha$ but the same holds for $\tilde{\alpha}$. This is because even though $\tilde{\alpha}$ is a nonlinear transformation from $\alpha$, it is still monotonic, hence the Spearman Rank correlation remains the same.}
\be
 p(b_1|\,\tilde{\alpha},c,m) \simeq p(b_1|\,\tilde{\alpha},m).
 \label{lastpaperresult}
\ee
Thus, the assembly bias trends $c\leftrightarrow b_1$ reflect the two fundamental correlations $c\leftrightarrow\tilde{\alpha}$ and $b_1\leftrightarrow\tilde{\alpha}$. 
This also implies that, given our formalism for modeling $b_{1}(\tilde{\alpha},m)$, we should also be able to predict $b_{1}(c,m)$, provided we know the correlation coefficient $\rho$ between $\tilde\alpha$ and $c$. We pursue this idea in this section by developing a bivariate model of halo assembly bias.

\subsection{Bivariate Lognormal Model}
\label{bivariatelognormal}
Considering $b_{1}$ as a stochastic property for every halo, we can think of the mean bias at fixed halo mass as the expectation value
\be
\avg{b_1|m} = \int\der b_1\,p(b_1|m)\,b_1\, \equiv b_1(m).
\ee
Similarly, conditional averages of $b_1$ can be expressed in terms of appropriate probability distributions as follows,
\begin{align}
\avg{b_1|c,m} &= \int\der b_1\,p(b_1|c,m)\,b_1\notag\\
&=\int\der b_1\int\der\tilde\alpha\,p(b_1|\tilde\alpha,c,m)\,p(\tilde\alpha|c,m)\,b_1\notag\\
&= \int\der\tilde\alpha\,\avg{b_1|\tilde\alpha,m}\,p(\tilde\alpha|c,m)\,,
  \label{int1}
\end{align}
where we marginalized over $\tilde\alpha$ in the second line and assumed the conditional independence of $b_1$ on $c$ at fixed $\tilde\alpha$ in the last line (see equation~\ref{lastpaperresult}).
This simplifies the expression since we can now replace 
$\avg{b_{1}|\tilde{\alpha},m}$ as
\begin{align}
\avg{ b_{1}|\tilde{\alpha},m} &= 1 + b_{1}^{L}(m,\tilde{\alpha})g(z)^{-1}\,,
 \label{exp1}
\end{align}
where one obtains $b_{1}^{L}(m,\tilde{\alpha})$ from \eqn{eq:bias(m,s)}.
We can see that the $\tilde{\alpha}$ dependence occurs only in the Hermite polynomials, so we need to evaluate the following set of integrals
\be
 \avg{H_n(\tilde\alpha)|c,m} = \int H_{n}(\tilde{\alpha})p(\tilde{\alpha}|c,m) d\tilde{\alpha}\,,
\ee
So far, we have not discussed the distribution of the internal property $\rm c$. In the case where this distribution is standard normal, the above integral has an analytic solution,
\be
 \avg{H_n(\tilde\alpha)|c,m}= \int H_{n}(\tilde{\alpha})p(\tilde{\alpha}|c,m) d\tilde{\alpha} = \rho(m)^n H_{n}(c),
\ee
where $\rho(m)$ is the correlation coefficient between $c$ and $\tilde{\alpha}$ in the mass bin $m$ (see Appendix~\ref{appendix:hermiteintegral} for details).
Putting this back in equation~\eqref{int1} and \eqref{exp1} gives us 
\be
\begin{split}
 b_{1}(m,z,c)&\equiv  \avg{b_{1}|c,m,z}, \\
 &= b_{1}(m,z) + \mu_{1}^{L}(m,z) \rho(m,z) H_{1}(c)
  \\
 &\hspace{24pt}+\dfrac{1}{2}\Sigma_{1}^{L}(m,z) \rho^2(m,z) H_{2}(c)\,. \\
 \end{split}
 \label{eq:framework_extn_b1}
\ee
Note that by setting $\rho=1$ in the above equation, we can recover equation~\eqref{eq:bias(m,s)} as it should be in the case of $c=\tilde{\alpha}$.
Thus equation~\eqref{eq:framework_extn_b1} provides us with a continuous prediction for the dependence of bias on mass, redshift and any internal halo property that can be transformed to follow Gaussian distribution. Below, we will demonstrate this for halo concentration. 
\subsection{An Example: Halo Concentration}
\label{haloconc}
Halo concentration has been extensively used to describe halo assembly bias in the literature \citep{wechsler+06,jsm07,dwbs08,desjacques:2008a,abl08,fw10,shpl16}, although there are several other halo properties in which assembly bias manifests. Despite the large number of studies describing its assembly bias, there are relatively few attempts at accurately calibrating the effect (\citealp{wechsler+06}; \citealp{pp17}). Here, we provide an alternate calibration for the dependence of bias on halo concentration within the extended framework described in the previous sections. Halo concentration has an approximately Lognormal distribution, which makes it convenient for using its Gaussianized form as an example of the property $c$ in the bivariate Lognormal model introduced above. 

Denoting halo concentration by $c_{\rm 200b}=R_{\rm 200b}/r_{\rm s}$, where $r_{\rm s}$ is the scale radius of the NFW profile (\citealp*{nfw96}; \citealp*{nfw97}), we define the standardized variable $\tilde{c}_{200b}$ as 
\be      
 \tilde{c}_{200b} \equiv \dfrac{\rm \ln c_{200b}-\mu_{0}^{\prime}}{\sigma_0^{\prime}}
\label{eq:redefinec200b}
 \ee
where
\begin{align}
 \mu_{0}^{\prime} &\equiv \avg{\ln c_{200b}|m,\delta_L =0} \,,
\label{eq:mu0prime}\\
\sigma_0^{\prime2} &\equiv {\rm Var} (\ln c_{200b}|m,\delta_L=0) \,.
\label{eq:sigma0prime}
\end{align}
Previous work has hinted that the cause of concentration assembly bias is due to its association with the tidal environment. In the following, we show that the bivariate Lognormal model, which is based on this association, matches well with the simulations.

Note that this section gives just one example of the application of the bivariate model. There are secondary halo properties whose assembly bias has been demonstrated in the literature \citep{fw10} like velocity anisotropy $\beta$, which is near-Gaussian and halo spin $\lambda$, which is near-Lognormal. We can also use the halo properties $c_{x}/a_{x}$ and $c_{v}/a_{v}$, which are the ratios of the smallest to largest eigenvalues of the halo moment-of-inertia and velocity dispersion tensors, respectively, and are both near-Gaussian distributed. 
The dependence of bias on all of these halo properties can be calibrated in this formalism. We leave these for future work.

\subsubsection{Correlation Coefficient}
\label{seccorr}
\begin{figure}
    \centering    
    \includegraphics[width=\linewidth]{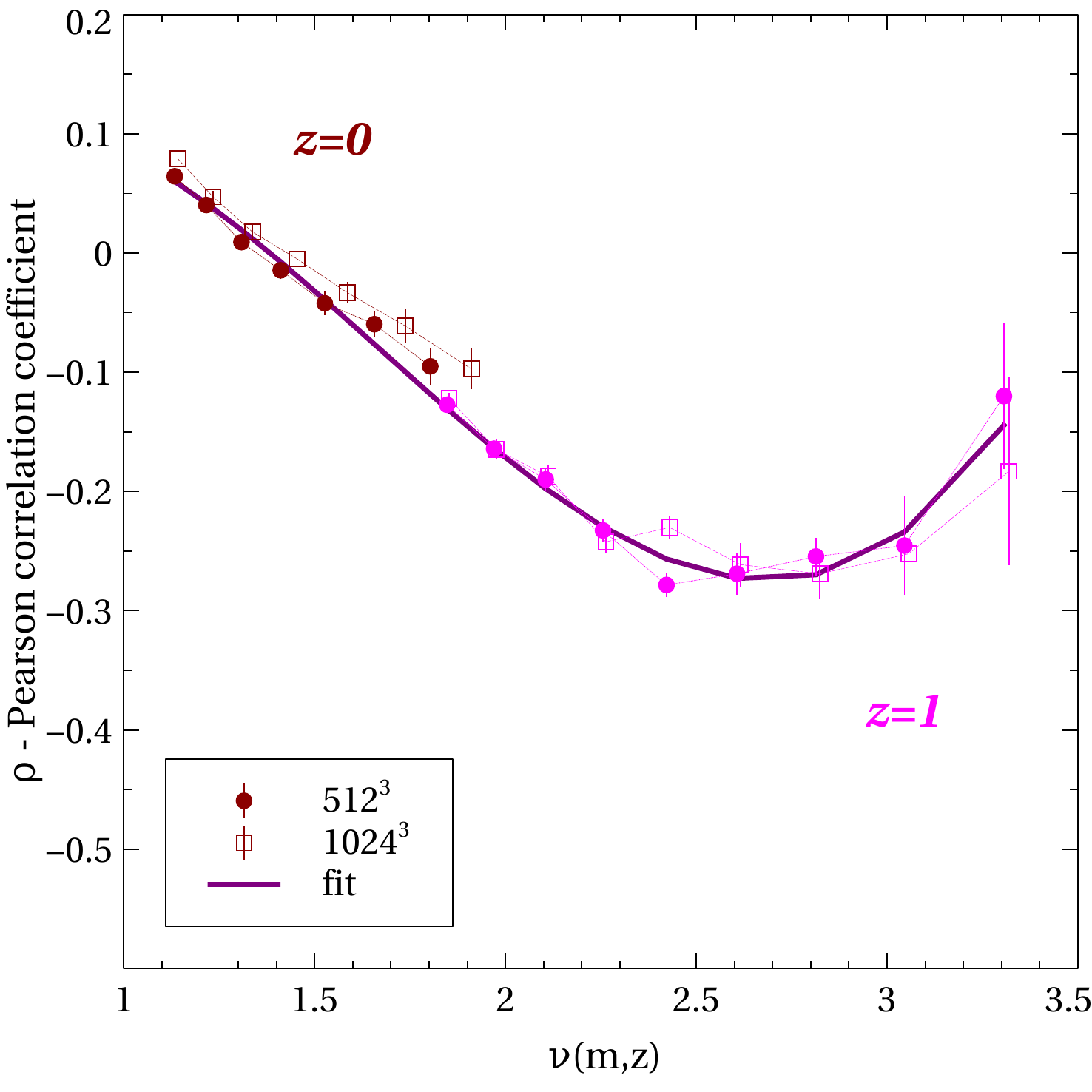} 
        \caption{Correlation $\rho$ between Gaussianized tidal anisotropy $\tilde{\alpha}$ (equation~\ref{eq:redefinealpha}) and halo concentration $c$ (equation~\ref{eq:redefinec200b}) using the first method 
         as a function of peak height $\nu$. The solid markers and filled markers show $\rho$ for default simulation and high-resolution simulation respectively. The best fit values and errors for the cubic fit (solid curve) are given in Table~\ref{table:rho}.} 
        \label{fig:correlation}
\end{figure}

\begin{table}
\centering
\begin{tabular}{cccccc}
 \hline 
 \hline 
 &&&&&$\chi^2$\\
&$\rho_{0}$&$\rho_{1}$&$\rho_{2}$&$\rho_{3}$&12 d.o.f \\ \hline 
value&-0.184&-0.247&0.106&0.092&17.35 \\ \hline 
std dev&0.003&0.011&0.011&0.017& \\ \hline 
corr $\rho_{0}$&1.000&0.367&-0.549&-0.394& \\ \hline 
corr $\rho_{1}$&-&1.000&-0.042&-0.794& \\ \hline 
corr $\rho_{2}$&-&-&1.000&0.555& \\ \hline 
 
 \hline \\ 
\end{tabular}
\caption{Best fit coefficients and covariance matrix of a cubic polynomial fit for $\rho$ as a function of pivoted peak height $\nu_p \equiv \nu-2.05$ : $\rho = \rho_{0}+\rho_{1}\nu_p+\rho_{2}\nu_p^2+\rho_{3}\nu_p^3$. The first row gives the least squares best fit values, the second row gives the standard deviation (square root of the diagonal elements of the covariance matrix). The last three rows give the correlation coefficients (elements of the covariance matrix $C_{ij}$ divided by $\sqrt{C_{ii}C_{jj}}$).}
\label{table:rho}
\end{table}

To describe the assembly bias with halo concentration, we require, in addition to $\mu_{1}^L$ and $\Sigma_1^L $ from Table~\ref{table:alpha}, knowledge about the correlation coefficient between tidal anisotropy and concentration. Here we have several options. The model mandates the use of Pearson's correlation coefficient. We can compute either the correlation coefficient of the Lognormal variables $\rho_{\rm LN}$ or their Gaussianized form $\rho$, both related to each other via the relation. 
\be
 \rho_{\rm LN} = \dfrac{e^{\rho \sigma_{\tilde{\alpha}} \sigma_{\tilde{c}_{200b}}} -1}{\sqrt{(e^{\sigma_{
 \tilde{\alpha}}^2}-1)(e^{\sigma_{
 \tilde{c}_{200b}}^2}-1)}}
 \label{corr_relation_equation}
\ee
where $\sigma_{\tilde{\alpha}}$ and $\sigma_{\tilde{c}_{200b}}$ are the standard deviation of Gaussianized tidal anisotropy $\tilde{\alpha}$ and concentration $\tilde{c}_{200b}$.

Details for obtaining \eqn{corr_relation_equation} are given in Appendix~\ref{corr_relation}.
However, when calculating Pearson's correlation coefficient for actual data, one needs to be wary that it is highly sensitive to outliers. The Spearman correlation coefficient is a good alternative which is robust against outliers, but its magnitude can differ from Pearson's correlation coefficient as required in \eqn{eq:framework_extn_b1}. 

We have identified three methods that we can use to compute the correlation coefficient $\rho$.
\begin{enumerate}
 \item First method: Compute Pearson's correlation coefficient $\rho_{\rm LN}$ between the Lognormal variables from the simulation and analytically obtain $\rho$ using \eqn{corr_relation_equation}.
 \item Second method: Gaussianize the tidal anisotropy and halo concentration and then obtain their correlation coefficient $\rho$. 
 \item Third method: Compute the Spearman correlation coefficient between the two variables.
\end{enumerate}

\begin{figure}
 \includegraphics[width=0.95\linewidth]{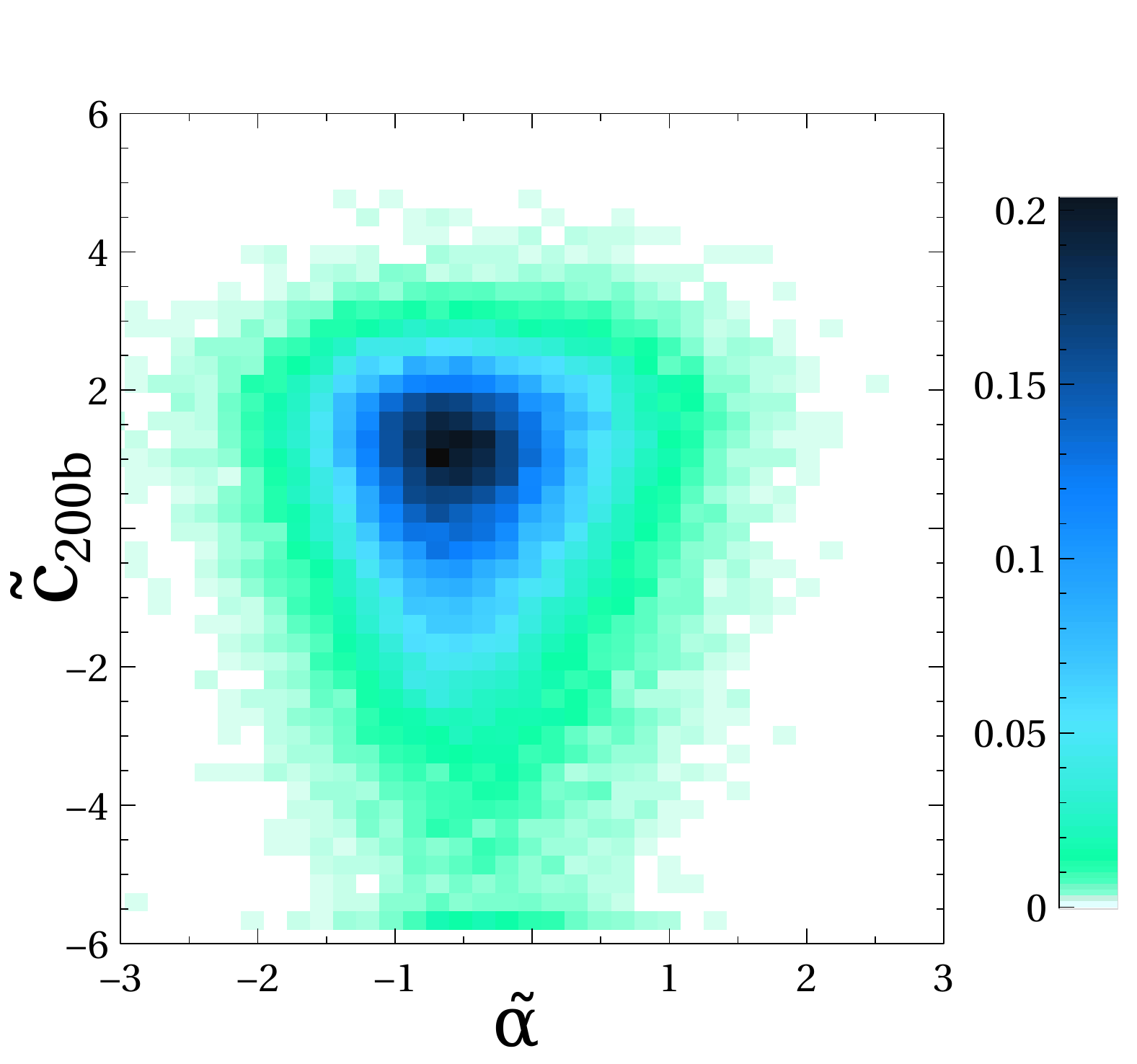}\\
 \caption{Bivariate distribution of tidal anisotropy $\tilde{\alpha}$ and Gaussianized halo concentration $\tilde{c}_{200b}$ in the mass range $1-1.5\times 10^{13}\Msun$. We can see that $\tilde{c}_{200b}$ has non-Gaussian outliers resulting in a negative tail.  }
 \label{distribution-c200b}
\end{figure}

Though all the methods should give similar results, they give slightly different values due to non-Gaussianities/outliers in the distribution of $\tilde{c}_{200b}$ and $\tilde{\alpha}$. The distribution of Gaussianized halo concentration $\tilde{c}_{200b}$ particularly has a negative skew as well as negative outliers, as can be seen in Figure~\ref{distribution-c200b}. Thus the already weak correlations become increasingly difficult to calculate accurately. We need to identify the method robust to these issues. After the detailed analysis done in Appendix~\ref{appendix:outliers}, we choose to work with the first method because we see that Pearson's correlation coefficient for Lognormal variables is more robust to negative outliers and downweights their influence in the calculation of the correlation coefficient. 

Figure~\ref{fig:correlation} shows $\rho$ as obtained from the first method as a function of peak height. We see an overall preference of the high mass haloes with high concentration to be in isotropic tidal environments. This trend reverses for low mass haloes; the highly concentrated haloes preferentially populate anisotropic tidal environments (more discussion below). A more detailed study at every mass range reveals non-monotonic relation between halo concentration and tidal anisotropy \citep[see Figure 12 of ][for more details]{phs18a}, however, we do not need to work at this level of detail here. This is because the correlation coefficient $\rho$ decisively captures the assembly bias signal associated with halo concentration at every mass range. This can also be seen reflected in the peak height of zero correlation ($\nu\sim 1.3$), which is identical to the peak height where the assembly bias signal inverts (compare with Figure~\ref{fig:ab_c200b}). These trends are in agreement with previous studies that look at environmental dependence of halo concentration at fixed mass and redshift \citep{wechsler+06,dwbs08,cs13}.

We choose to fit a third degree polynomial to this relation  after an analysis with Akaike information criteria with correction \citep{akaike1974,sugiura1978} for polynomials of various degree. The best  coefficients and covariance matrix are shown in Table~\ref{table:rho}.
We do all the subsequent analysis with this functional form. We have repeated the entire analysis using the other methods and find qualitatively similar results, although quantitative details differ. 

\subsubsection{Comparison with simulations}

\begin{figure}
    \centering
        \includegraphics[width=0.98\linewidth]{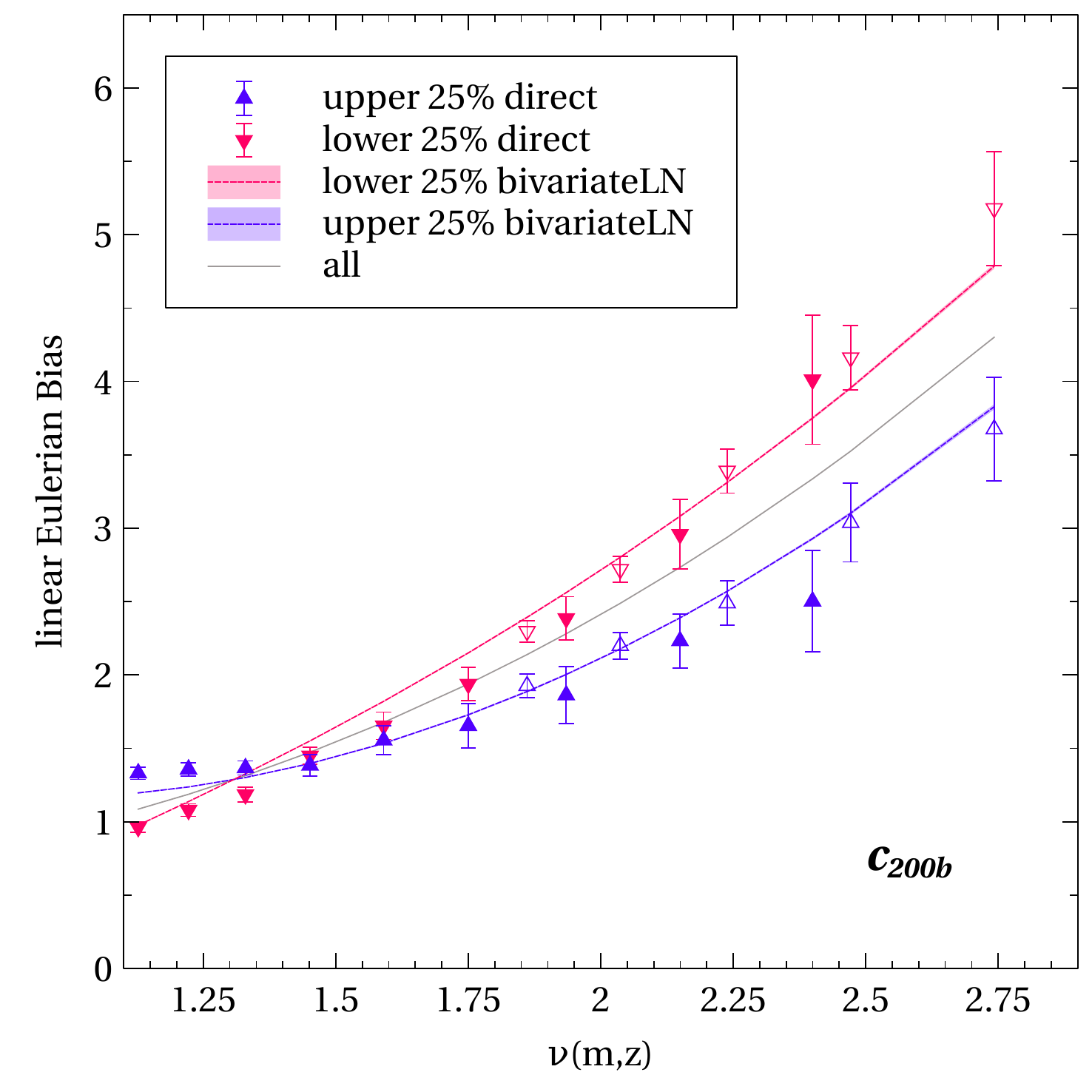} 
\caption{Linear halo bias $b_1$ 
as a function of peak height $\nu$ for upper and lower quartiles  of $c_{200b}$ population.
The data points with error bars are obtained from simulations. The two solid curves are obtained by taking $c=\tilde{c}_{200b}$ in the bivariate Lognormal model (see Sec~\ref{bivariatelognormal}). The error band around the solid curve is obtained from sampling
$\mu_{n}$ and $\Sigma_{n}^L$ from the covariance matrix in Table~\ref{table:alpha} and sampling $\rho$ from Table~\ref{table:rho} to obtain the standard deviation of $b_{1}(m,\tilde{c}_{200b})$ in the same way as described in Figure~\ref{fig:ab_alpha}.
}
\label{fig:ab_c200b}
\end{figure}

\noindent
We separately perform SU calculations as described in Section~\ref{sec:computeb1} for obtaining the peak-background split bias of 
halo populations for upper and lower quartiles of $c_{200b}$. In Figure~\ref{fig:ab_c200b},
the two sets of points with error bars show the bias for the upper and lower quartiles of $c_{200b}$. We compare this with the bivariate Lognormal model plotted as solid curves by averaging \eqn{eq:framework_extn_b1} above $\tilde{c}_{200b}>0.675$ and below $\tilde{c}_{200b}<-0.675$ for the upper and lower quartiles of halo concentration respectively. 
Error bands are obtained in the same manner as before in the case of assembly bias in $\alpha$, the covariance matrix from Table~\ref{table:alpha} is used to construct a trivariate Gaussian distribution and the coefficients $\mu_{1}^L$ and $\Sigma_1^L $ are sampled 300 times to obtain convergent error values. $b_1^L(m,\tilde{c}_{200b})$ is computed each of these times. The standard deviation of the above sample of $b_1^L(m,\tilde{c}_{200b})$ is plotted as a band around the Lognormal model.

\subsection{Can the model predict quadratic assembly bias?} 
\label{subsec:predictb2}

\begin{figure}
    \centering
        \includegraphics[width=0.95\linewidth]{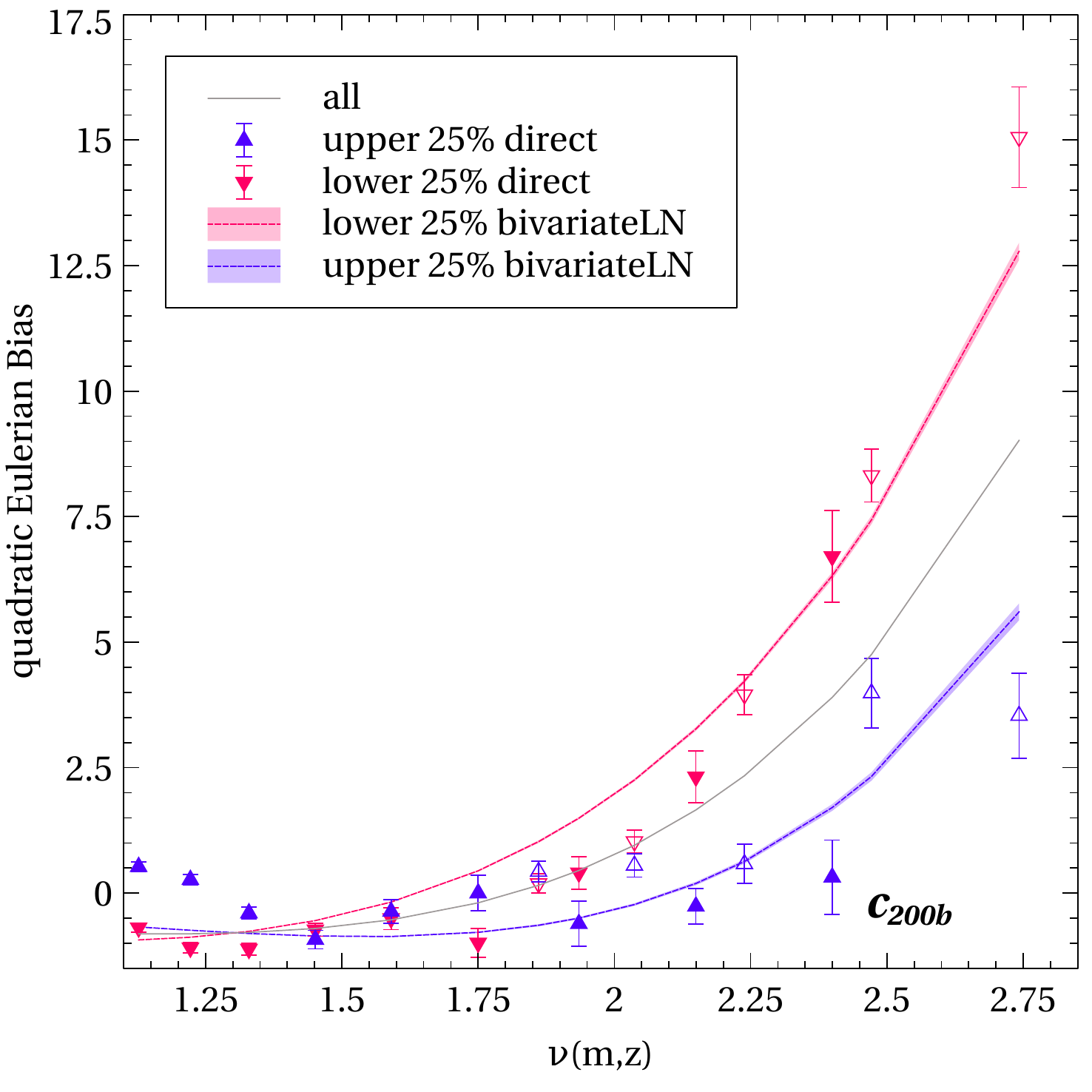} 
\caption{Quadratic halo bias $b_{2}$ 
as a function of peak height $\nu$ for upper and lower quartiles of $c_{200b}$ population. 
The points and curves are formatted similar to Figure~\ref{fig:ab_c200b} and show a comparison of simulation measurements with the bivariate Lognormal model for $b_{2}(m,\tilde{c}_{200b})$. }
\label{fig:ab2_c200b}
\end{figure}
So far, in Section~\ref{haloconc}, we have used the conditional independence of linear bias $b_1$ and a halo property in fixed tidal environments
to predict the linear assembly bias with the property $c$. \citet{rphs19} showed this by treating linear halo bias as a halo-centric property and computing correlation coefficients with other halo-centric quantities. In principle, one could verify the same for quadratic bias by measuring the bispectrum and calculating an analogous `halo-by-halo quadratic bias'.
Instead, here we assume the conditional independence of $b_2$ and an internal property of the halo, i.e.,
 \be
 \avg{b_2|\tilde{\alpha},c,m} = \avg{b_2|\tilde{\alpha},m}\,,
 \label{lastpaperresult_assumption}
\ee
using which we model the quadratic assembly bias with halo property $c$. 
The resulting dependence of $b_2$ on halo mass and halo property $c$ can be written, analogous to \eqn{eq:framework_extn_b1}, as
\be
\begin{split}
 b_{2}(m,c,z) = b_{2}+\{\mu_{2}^L+2\mu_{1}^L(b_{1}-1)+\frac{8}{21}\mu_{1}^L\} \rho H_{1}(c)
 \\
 +\{(\mu_{1}^L)^2+\Sigma_{1}^L(b_{1}-1)+\frac{1}{2}\Sigma_{2}^L+\frac{4}{21}\Sigma_{1}^L\}\rho^2 H_{2}(c) \\
 +\mu_{1}^L\Sigma_{1}^L \rho^3 H_{3}(c)+\frac{1}{4}(\Sigma^L_{1})^2 \rho^4 H_{4}(c)\,,
 \end{split}
 \label{eq:framework_extn_b2}
\ee
For brevity, we have suppressed the mass and redshift dependence on all terms on the right side of the equation above except the Hermite polynomials, which only have c dependence.
We test the accuracy of the above equation in Figure~\ref{fig:ab2_c200b}. Although the model qualitatively describes the simulation points, the overall agreement is poor at low masses. This could be due to the systematic error in the measurement of second-order terms. It could also be that the assumption of conditional independence in \eqn{lastpaperresult_assumption} breaks down for higher-order non-linear bias coefficients at these mass scales. This is not perhaps unexpected since the low mass haloes are a mix of two kinds of populations in contrasting environments, making their trends complicated. One subpopulation of haloes in isotropic environment behave like ‘standard’
peaks theory/excursion set haloes, and their halo concentration is negatively correlated with the large scale density environment. In contrast, the other subpopulation lives in a highly anisotropic environment, initially set to become high mass haloes, but get tidally truncated by redirected mass flow to filaments, and their halo concentration is positively correlated with the environment (\citealp{hahn+09}; \citealp{phs18a}). A fuller exploration of these effects would be possible using direct measurements of the halo bispectrum in different tidal environments, an exercise we leave to future work.

\section{Summary}

\label{sec:summary}
Halo assembly bias is a potential source of systematic uncertainty for cosmological inference from upcoming large-volume galaxy surveys, as well as being a possible channel for enhancing our understanding of galaxy formation and evolution. Our aim in this work has been to develop accurate calibrations of the dependence of halo bias on one of the primary `beyond halo mass' variables responsible for assembly bias, namely, the tidal anisotropy $\alpha$ of the local cosmic web environment of haloes. We used the Separate Universe (SU) technique to calibrate the dependence of linear and quadratic bias $b_1$ and $b_2$, respectively, on halo mass, redshift and $\alpha$. We also showed, using the example of halo concentration, that it is possible to make use of this calibration on web environment to further calibrate the dependence of bias on other secondary properties. Our results can be summarized as follows:
\begin{itemize}
    \item The tidal anisotropy $\alpha$ has a nearly Lognormal distribution over the entire range of peak height that we studied $1.1\lesssim \nu \lesssim 3.4$, summarized in Table~\ref{table:alpha_mean_variance}.  
    \item  We first used the SU approach to numerically calculate $b_1(\alpha,\nu)$ (Figure~\ref{fig:ab_alpha}) and $b_2(\alpha,\nu)$ (Figure~\ref{fig:ab2_alpha}) in  quartiles of $\alpha$ and bins of peak height $\nu(m,z)$. 
    This is the first reported detection  of quadratic assembly bias with respect to the tidal environment of the halo\footnote{Every $\delta_L$ box has a maximum mass, which can be probed before getting affected by small sample size, as we are limited by the finite comoving volume of our simulations. We can only probe masses upto $\nu\lesssim 2.8$ in our least dense box ($\delta_{L}=-0.7$), which is used in all analyses with SU approach. In the analysis involving only $\delta_L=0$ we can probe upto $\nu\lesssim 3.4$.}.
    \item We also analytically calibrated, with very high precision, the relations $b_1(\alpha,\nu)$ and $b_2(\alpha,\nu)$ as \emph{continuous functions} of $\alpha$ (i.e., without binning) using the framework developed in \citet{pp17} (see Section~\ref{sec:lnmodel}) which exploits the near-Lognormal distribution of $\alpha$, combined with fitting functions $b_1(\nu)$ and $b_2(\nu)$ from the literature for the all-halo results. These results are summarized in equations~\eqref{eq:bias(m,s)}-\eqref{eq:bias2(m,s)}, Table~\ref{table:alpha} and Figures \ref{fig:ab_alpha}, \ref{fig:b1fixedalpha} and \ref{fig:ab2_alpha}, with a comparison to the $\alpha$-dependence of linear bias directly measured in simulations shown in Figure~\ref{fig:ab_directmmn}. 
   \item Using the conditional independence of large-scale bias on secondary halo properties at fixed $\alpha$ \citep{rphs19}, we then extended this analytic framework to
   accommodate the dependence of bias on another secondary property, whose distribution has or can be monotonically transformed to have a Gaussian form (Section~\ref{bivariatelognormal}). We demonstrated this technique for the case of halo concentration $c_{200b}$ by calibrating the conditional distribution $p(c_{200b}|\alpha,\nu)$ (Figure~\ref{fig:correlation} and Table~\ref{table:rho}). We reproduce the known dependence of $b_1(c_{200b},\nu)$ accurately over our entire dynamic range (Figure~\ref{fig:ab_c200b}), while $b_2(c_{200b},\nu)$ departs from previous results at low $\nu$. We discussed possible reasons for the latter discrepancy in Section~\ref{subsec:predictb2}.
    \end{itemize}
    
    Our calibrations of $b_1$ and $b_2$ can potentially be useful in a number of areas:
    \enumerate{\item  Self-calibrating cluster surveys which constrain cosmological parameters as well as mass-observable relations (\citealp{mm04}; \citealp{wrw08}; \citealp{cooaul20}; \citealp{nds20}). 
    \item Redshift space distortion (RSD) modeling to constrain cosmic acceleration physics:  This can be done by incorporating correlations between large scale bias and velocity dispersion into RSD modeling which can potentially constrain cosmological parameters sensitive to the nature of gravity.
    \item The calibration of $b_1,b_2$ on tidal anisotropy and mass provides a possibility to improve models which use three-point statistics like the bispectrum to constrain primordial non-Gaussianities (\citealp{jk09}; \citealp{kllrbv18}; \citealp{gv20}).
    \item Analytical forecasts for multi-tracer analyses that require samples with widely different bias parameters \citep{mCds09,fcsm15}.
    \item Halo occupation distribution modeling to incorporate assembly bias in mock catalogs, potentially for several secondary properties in addition to $\alpha$ and halo concentration discussed here \citep[see, e.g.,][for recent work along these lines]{xzc20}.}
    
We will return to these topics in future work.

\section*{Acknowledgments}
We thank Ravi Sheth and Oliver Hahn for useful discussions and the anonymous referee for a helpful report. We thank the Munich Institute for Astro- and Particle
Physics (MIAPP) and the organisers of the programme on
Dynamics of Large-Scale Structure (July 2019) for their hospitality while part of this work was completed.
The research
of AP is supported by the Associateship Scheme of ICTP,
Trieste and the Ramanujan Fellowship awarded by the Department of Science and Technology, Government of India. We gratefully acknowledge the use of high performance computing facilities at IUCAA, Pune.\footnote{\href{http://hpc.iucaa.in}{http://hpc.iucaa.in}}

\section*{Data Availability}
No new data were generated in support of this research. The simulations used in this work are available from the authors upon reasonable request. The code for implementing the analytical framework is also supplied online with sample demonstrations \href{https://github.com/rsujatha/CAB}{\faGithub}.
\bibliography{masterRef}
\appendix
\section{Sensitivity of Correlation Coefficients to outliers}
\label{appendix:outliers}

\begin{table*}
\centering
 \begin{tabular}{llllll}
 \hline 
 \hline  
Outliers ($\tilde{\alpha}$,c)&Population&True $\rho$&Pearson's $\rho$&Pearson's $\rho$&Spearman $\rho$ \\ 
&fraction&&(first method)&(second method)&(third method) \\ 
 \hline
($\tilde{\alpha} \in \mathcal{N}$(0,1),-30) &0.16 \%   & 0.02&  0.0201 & 0.0129 &  0.0192   \\ 
($\tilde{\alpha} \in \mathcal{N}$(0,1),-30)  &3.33\%  & 0.02& 0.0191& 0.0034 & 0.0177    \\ 
($\tilde{\alpha} \in \mathcal{N}$(0,1),-40)&  0.16\%  & 0.02&0.0200 & 0.0107 & 0.0192    \\
($\tilde{\alpha} \in \mathcal{N}$(0,1),-40) & 3.33\%  &  0.02&0.0201 & 0.0021 & 0.0177  \\ 
($\tilde{\alpha}\in \mathcal{N}(0,1),-2 e^{\tilde{2\alpha}}$)&0.16\% &0.02 & 0.0179&-0.0521&0.0094 \\
( -40    , -40 )&0.16\%&-0.5&-0.49&0.59&-0.47\\
\hline 
 \hline \\ 
\end{tabular}
\caption{Robustness of correlation coefficients to various  kinds of outliers. A sample of 600,000 is made by first sampling bivariate normal distribution with mean and variance of both variables 0 and 1 respectively. The true correlation of the sample is given in the `True $\rho$' column. Outliers are added to this sample as per the table and the correlation coefficients recalculated to check their sensitivity. See Appendix~\ref{appendix:outliers} for a description of the method.}
\label{table:outliers}
\end{table*}

There are three ways to measure the correlation coefficient between two variables $\tilde{\alpha}$ and $\tilde{c}_{200b}$ as described in the main text in Section~\ref{seccorr}.  The first method computes  Pearson's correlation coefficient between the halo tidal anisotropy and concentration. These are Lognormal variables, and we use \eqn{corr_relation_equation} to obtain the correlation coefficient between their Gaussianized forms. The second method converts them to Gaussianized form first and computes  Pearson's correlation coefficient. The third method computes the Spearman rank correlation coefficient between the two variables.

In this section, we want to select the best method of estimating the correlation coefficient from the three listed above. We do this by considering a toy exercise using a mock sample of 600,000 ``haloes". 
Each halo is assigned two properties ($\tilde{\alpha}$ and $\tilde{c}_{200b}$), which are distributed as a bivariate Gaussian. Each of these properties has zero mean and unit variance, with the correlation coefficient 0.02. This correlation coefficient is chosen as it is one among the weakest correlations seen in our simulations between $\tilde{\alpha}$ and $\tilde{c}_{200b}$, hence most sensitive to the choice of method.

To this population, we add various outliers, as shown in the first column of Table~\ref{table:outliers}. These outliers are represented by tuples describing the two properties $\tilde{\alpha}$ and $\tilde{c}_{200b}$ of the halo. 
Recall that our goal is to choose the method least sensitive to the presence of negative outliers as this is especially the case in halo concentration $\tilde{c}_{200b}$ though not so much in $\tilde{\alpha}$ distribution (see Figure~\ref{distribution-c200b}). So we construct toy examples where the outlier haloes have negative skewness in $\tilde{c}_{200b}$, while $\tilde{\alpha}$ is chosen from a standard Gaussian so as to preserve its marginal distribution. 
The second column shows the percentage of the population that comprises of the outliers. The third column shows the true correlation (excluding the outliers) of the population. The last three columns show how the correlation coefficient deviates from the true correlation for the three computing methods. 

For example, in the first row of the table, ($\tilde{\alpha} \in \mathcal{N}$(0,1),-30) means that the outliers have $\tilde{c}_{200b}=-30$ and $\tilde{\alpha}$ is drawn from standard Normal distribution $\mathcal{N}$(0,1) 
and they comprise $0.16\%$ of the total population. Out of the three methods, the first method is the most robust and closest to the true correlation while the second method is most sensitive to outliers. 
In fact, in all other examples, the first method is the most robust to the presence of outliers.

The last example, which is an extreme case of large negative outliers in both $\tilde{\alpha}$ and $\tilde{c}_{200b}$, is used to demonstrate why the first method works better than the rest in the presence of a small population of highly negative outliers. We can see that a true negative correlation of $-0.5$ can turn to an even higher positive correlation of $0.59$ when calculated using Pearson's second method. 
To understand why the first method works better, let us reconstruct the two Lognormal variables $\alpha = \exp(\tilde{\alpha}\sigma_{0}+\mu_{0})$ and $c_{200b} = \exp(\tilde{c}_{200b} \sigma_{0}^{\prime}+\mu_0^{\prime})$ where $\mu_{0}$, $\sigma_{0}$, $\mu_{0}^{\prime}$, $\sigma_{0}^{\prime}$ are as defined in \eqns{eq:mu0}, \eqref{eq:sigma0}, \eqref{eq:mu0prime} and \eqref{eq:sigma0prime}. While computing  Pearson's correlation coefficient in a simulation with $N$ haloes, the presence of $\mu_{0}$ and $\mu_{0}^{\prime}$ would cancel to give

\be
\rho_{\rm LN} =\frac{\sum\limits_{i} \frac{ e^{\tilde{\alpha}^{i}\sigma_{0}+\tilde{c}_{200b }^i\sigma_{0}^{\prime}}}{N}-\sum\limits_{j}\frac{e^{\tilde{\alpha}^{j}\sigma_{0}}}{N}\sum\limits_{k}\frac{e^{\tilde{c}^{k}_{200b}\sigma_{0}^{\prime}}}{N}}{S(\sigma_{0},\tilde{\alpha})S(\sigma_{0}^{\prime},\tilde{c}_{200b})}\\
\ee
where $S(\sigma_{0},\tilde{\alpha})$ is given by

\be
S(\sigma_{0},\tilde{\alpha}) = \sqrt{\sum\limits_{j} \frac{e^{2\tilde{\alpha}^{j}\sigma_{0}}}{N}-(\sum\limits_{j^{\prime}}\frac{e^{\tilde{\alpha}^{j^{\prime}}\sigma_{0}}}{N})^2} \notag
\ee
where the summation is over all the haloes. When written in this form, it becomes easy to see how negative outliers will be exponentiated and thus contribute negligibly to the above summations, leaving the correlation coefficient robust to these highly negative outliers. However, this method need not be restricted to be used for suppressing outliers of a negatively skewed Gaussian distribution; the contribution from a positive skew of a near-Gaussian variable ($a$) can also be suppressed by this method with additional steps: transform the variable $a\to-a$  before applying the method and transform the correlation coefficient $\rho\to-\rho$ after applying the method.

 We do not forget that in the attempt to conform to the Gaussian distribution that the model mandates, we have ignored a fraction of haloes having unusually low concentration, a population that could be physically interesting. One could, in principle, use Gaussian mixtures to factor in the tail as has been done in \citet{neto+07}, where the distribution is a sum of a larger Gaussian and a smaller one with smaller mean and larger variance. We leave such explorations for future work.

\section{Correlation coefficients (Log)normal variables}
\label{corr_relation}
Let $Z_{i}$ be a random variable with Lognormal distribution, 
\begin{align}
\notag
Z_{i} &= \exp{X_{i}} ,\quad X_{i} \sim \mathcal{N}(\mu_{i},\sigma^{2})\,.
\notag
\end{align}
The mean of $Z_i$ can be written as
\begin{align}
\notag
 E(Z_{i})
 &=\exp(\mu_{i}) \exp(\sigma_{i}^2/2)\,.
\end{align}
This can be deduced from the one variable equivalent expression of the Moment generating function $M_{X}(t)$ for multivariate correlated variables,
\be
M_{X}(t^TX) = \exp{t^T\mu + \frac{1}{2}t^T \Sigma t}
\ee
where $X,t,\mu$ are $n$ dimensional vectors and $\Sigma$ is the covariance matrix. Then the expectation value of $Z_{i}^2$ can also be written as
\begin{align}
\notag
 E(Z_{i}^2)&= \avg{\exp{(2X_{i})}}\\
 \notag
 &= \exp(2\mu_{i})\exp(2\sigma_{i}^2)
\end{align}
Hence we find the variance of $Z_{i}$ to be
\begin{align}
\notag
 {\rm Var}(Z_{i}) &= E(Z_{i}^2) - E(Z_{i})^2\\
 \notag
 &=\exp(2 \mu_{i}+2\sigma_{i}^2)-\exp(2 \mu_{i}+ \sigma_{i}^2)\\
 \notag
 &=\exp(2 \mu_{i})\exp\sigma_{i}^2(\exp(\sigma_{i}^2)-1)
\end{align}
Now consider two Lognormal variables $Z_{i}$ and $Z_{j}$ with correlation coefficeint $\rho$. The expectation value of their product is  
\begin{align}
\notag
 E(Z_{i}Z_{j}) &=\avg{\exp(X_{i}+X_{j})}\\
 \notag
 &=M_{\{X_{i},X_{j}\}}(t)|_{t=(1,1)}\\
 \notag
 &=\exp(\mu_{i}+\mu_{j})\exp \frac{1}{2}(\sigma_{i}^2+\sigma_{j}^2+2\rho \sigma_{i}\sigma_{j})
\end{align}
Now we can find the correlation coefficient $\rho_{\rm LN}$ between two Lognormal variables $Z_{i}$ and $Z_{j}$ to be
\begin{align}
\notag
 \rho_{\rm LN} &= \rho_{Z_{i}Z_{j}} = \dfrac{E(Z_{i}Z_{j}) - E(Z_{i})E(Z_{j})}{\sqrt{{\rm Var}( Z_{i}){\rm Var}( Z_{j})}}\\
 \notag
 &= \dfrac{e^{\mu_{i}+\mu_{j}}e^{1/2(\sigma_{i}^2+\sigma_{j}^2+2\rho \sigma_{i}\sigma_{j})}- e^{\mu_{i}+\mu_{j}}e^{1/2(\sigma_{i}^2+\sigma_{j}^2)}}{\sqrt{e^{2\mu_{i}+2\mu_{j}}e^{\sigma_{i}^2+\sigma_{j}^2}(e^{\sigma_{i}^2}-1)(e^{\sigma_{j}^2}-1)}}\\
 \therefore \rho_{\rm LN}&=\dfrac{e^{\rho \sigma_{i}\sigma_{j}}-1}{\sqrt{(e^{\sigma_{i}^2}-1)(e^{\sigma_{j}^2}-1)}}\,.
 \label{LNGauss_reln}
\end{align}

\section{Hermite Polynomial integral}
\label{appendix:hermiteintegral}
The probabilist's Hermite polynomials are defined by 
\be
p(s)H_n(s) = \left(-\der/\der s\right)^np(s) = \int\frac{\der k}{2\pi}\e{iks}(-ik)^n\e{-k^2/2}\,,
\label{eq:hermite}
\ee
where $p(s)=\e{-s^2/2}/\sqrt{2\pi}$ is the probability density function of a standard normal deviate. All integrals range from $-\infty$ to $\infty$ over the respective variable.

If both $\tilde{\alpha}$ and $\rm c$ are standard normal deviates with correlation coefficient $\rho$, then we have
\be
p(c|\tilde\alpha) = \int\frac{\der k}{2\pi}\e{ik(c-\rho\tilde\alpha)}\e{-k^2(1-\rho^2)/2}\,,
\ee
and we can write
\begin{align}
\avg{H_n(\tilde\alpha)|c} &= \int\der\tilde\alpha \,p(\tilde\alpha|c)H_n(\tilde\alpha)\notag\\
&= \frac{1}{p(c)}\int\der\tilde\alpha\,p(\tilde\alpha)p(c|\tilde\alpha)H_n(\tilde\alpha)\notag\\
&= \frac{1}{p(c)}\int\der\tilde\alpha\,p(c|\tilde\alpha)\left(-\frac{\p}{\p\tilde\alpha}\right)^n p(\tilde\alpha)\notag\\
&= \frac{1}{p(c)}\int\der\tilde\alpha\,p(\tilde\alpha)\left(\frac{\p}{\p\tilde\alpha}\right)^np(c|\tilde\alpha)\notag\\
&= \frac{\rho^n}{p(c)}\int\der\tilde\alpha\,p(\tilde\alpha)\int\frac{\der k}{2\pi}\e{ik(c-\rho\tilde\alpha)}(-ik)^n\e{-k^2(1-\rho^2)/2}\notag\\
&= \frac{\rho^n}{p(c)}\int\frac{\der k}{2\pi}\e{ikc}\avg{\e{-ik\rho\tilde\alpha}}(-ik)^n\e{-k^2(1-\rho^2)/2}\notag\\
&= \frac{\rho^n}{p(c)}\int\frac{\der k}{2\pi}\e{ikc}\e{-k^2\rho^2/2}(-ik)^n\e{-k^2(1-\rho^2)/2}\notag\\
&= \frac{\rho^n}{p(c)}\int\frac{\der k}{2\pi}\e{ikc}(-ik)^n\e{-k^2/2}\notag\\
&= \frac{\rho^n}{p(c)} H_n(c)\,p(c)\notag\\
\therefore \avg{H_n(\tilde\alpha)|c} &= \rho^n H_n(c)\,.
\end{align}
\label{lastpage}
\end{document}